\documentclass{article}
\usepackage[T1]{fontenc}
\usepackage[english]{babel}

\makeatletter
\let\@fnsymbol\@arabic
\makeatother

\usepackage{amssymb}
\usepackage{float}
\usepackage{subcaption}  % For creating subfigures
\usepackage{amsmath}
 \usepackage{amsthm}

\theoremstyle{plain}

\usepackage{tikz}
\usetikzlibrary{calc,positioning,arrows,arrows.meta,shapes.geometric}
\usetikzlibrary{arrows.meta,positioning,fit,backgrounds}

\usepackage{amsmath,amssymb,amsthm}
\usepackage{graphicx}
\usepackage[colorlinks=true, allcolors=blue]{hyperref}
\usepackage{xcolor}
\usepackage{booktabs}
\usepackage{array}
\usepackage{tikz}
\usetikzlibrary{calc,positioning,arrows,arrows.meta,shapes.geometric}
\usepackage{xurl}

\DeclareMathOperator{\swish}{swish}
\newcommand{\dR}{\mathbb{R}}

\author{Karim Cherifi\footnotemark[1]
\, \and Achraf El Messaoudi\footnotemark[1]\, \and Hannes Gernandt\footnotemark[2]\, \and Marco Roschkowski\footnotemark[2]}

\title{Nonlinear port-Hamiltonian system identification from input-state-output data (ISO-pHNN)} %% Article title

\begin{document}


%A. El Massaoudi, H. Gernandt, M. Roschkowski} %% Author name


%\affiliation[l2]{organization={Sorbonne University},%Department and Organization
 %           addressline={4 place Jussieu}, 
  %          city={Paris},
  %          postcode={75005}, 
  %          country={France}}

\renewcommand*{\thefootnote}{\fnsymbol{footnote}}
\maketitle

%% Abstract
\begin{abstract}
In this paper, we introduce a framework called ISO-pHNN for identifying nonlinear port-Hamiltonian systems using input-state-output data. The framework utilizes neural networks' universal approximation capacity to effectively represent complex dynamics in a structured way.  We explore different architectures based on MLPs, KANs, and using prior information. The identification technique is validated through examples featuring nonlinearities in either the skew-symmetric terms, the dissipative terms, or the Hamiltonian. We show that incorporating a port-Hamiltonian structure does not lower the accuracy and that using additional prior information improves long-term predictions.
\end{abstract}

\noindent
{\bf MSC:} 93B30,  93B15, 93C10,	68T07, 93B99 

\footnotetext[1]{
Université Marie et Louis Pasteur, SUPMICROTECH, CNRS, institut FEMTO-ST, Besançon, F-25000, France. \texttt{\{karim.cherifi,achraf.elmessaoudi\}@femto-st.fr}.}

\footnotetext[2]{
Institute for Mathematical Modeling, Analysis and Computational Mathematics, University of Wuppertal, Gau\ss stra\ss e 20, 42119 Wuppertal, Germany. \texttt{\{gernandt,roschkowski\}@uni-wuppertal.de}.}

\section{Introduction}
Physics-based models are widely used for simulating and controlling dynamical systems because of their interpretability and grounding in fundamental laws. However, deriving such models can be challenging for complex or proprietary systems. In this work, we focus on the identification of port-Hamiltonian (pH) models from input–state–output trajectories, a scenario that arises in applications where state information is accessible either through measurements, observers, or dimensionality reduction. This approach enables the recovery of structure-preserving, interpretable models that are suitable for control design, stability analysis, and modular interconnection with other pH systems.

While purely data-driven methods have been widely applied to model complex systems, they often face issues with generalizability and reliability beyond their training scope \cite{GorH24}. Our approach leverages data-driven techniques to extract the underlying pH structure. This is particularly valuable when the system dynamics are known only through simulation environments, high-dimensional models, or legacy tools, where explicit equations may not be available. By combining data-driven learning with the inherent structure of pH systems, we obtain models that preserve physical properties such as energy conservation and interconnection rules.

This hybrid approach that combines physics and Machine Learning (ML) is increasingly being adopted as it allows us to take advantage of the accuracy and generalizability of physics-based models while exploiting the flexibility and speed of ML methods. This integration has been particularly successful in several areas, such as the enhancement of physical models with deep networks to forecast complex dynamics~\cite{Yin2021}, the development of neural ordinary differential equations (Neural-ODEs) \cite{Chen2018}, or the use of Physics-Informed Neural Networks (PINNs) \cite{Raissi2019} to solve forward and inverse problems for (partial) differential equations. When the solver is known, ML can be used to learn the residual or inductive biases, as seen in the Neural-ODE approach \cite{Chen2018}, which combines physics from learning and solver-based methods. 
In cases where everything must be learned end-to-end, approaches such as PhyDNet and MeshGraphNet have been developed to disentangle physical dynamics from unknown factors, enabling the learning of complex simulations \cite{Guen2020, Pfaff2020}.

Hamiltonian neural networks represent a novel method for integrating extra physical knowledge into the dynamics that are to be identified. These networks leverage the Hamiltonian framework of dynamics to shape the neural architecture~\cite{Greydanus2019, Toth2020, Finzi2020,SchSBP24}. Although these models perform better than non-physics based models, these Hamiltonian-based models inherently represent closed systems and they do not incorporate dissipation in the model. 

In this study, we examine an expansion of Hamiltonian modeling, known as port-Hamiltonian (pH) systems, characterized as systems of ordinary differential equations structured as follows
\begin{align}
\label{eq:phs}
\begin{split}
\dot{x}(t) &= \left(J(x(t)) - R(x(t))\right) \nabla H(x(t)) + B(x(t))u(t),\quad x(0)=x_0 \\
y(t) &= B(x(t))^\top \nabla H(x(t)), \quad t\geq 0,% + \left(N(x) + S(x)\right)u,
\end{split}
\end{align}
where $x(t) \in \mathcal{X} \subseteq \mathbb{R}^n$ is the state vector, $u(t) \in \mathcal{U} \subseteq \mathbb{R}^m$ is the input vector,  $y(t) \in \mathcal{Y} \subseteq \mathbb{R}^m$ is the output vector and $\mathcal{X}$, $\mathcal{U}$, $\mathcal{Y}$ are some compact sets. When we need to indicate the dependency of the solutions $x$ of \eqref{eq:phs} on the initial value $x_0$ and the input $u$ then we write $x(t;x_0,u)$ and $y(t;x_0,u)$ for the corresponding output. We will from now on omit the time dependence of $x = x(t), u=u(t), y=y(t)$ to simplify the notation.
The Hamiltonian $H : \mathcal{X} \rightarrow \mathbb{R}$ is a continuously differentiable function that is assumed to be bounded from below. The matrices $J(x), R(x) \in \mathbb{R}^{n \times n}$, $B(x) \in \mathbb{R}^{n \times m}$,
satisfy the following properties:
\begin{equation}
J(x) = -J^\top(x),\quad  R(x)=R(x)^\top\geq 0, \quad  \forall x \in \mathcal{X}.
\label{eq:phsCond}
\end{equation}
 
Within this framework, the matrix function $J$ signifies the internal energy-conserving connections within the system, accounting for reversible interactions. Furthermore, the framework incorporates dissipation in the model when $R \neq 0$, attributable to factors such as resistance, friction losses, load losses resulting from electromagnetic effects in the windings of an electrical machine, and analogous effects. 

The class of pH systems is becoming increasingly popular because it offers robust models that are inherently passive and can guarantee the stability of equilibria, see \cite{MehU23,SchJ14,CheGH23} for an overview. Since the interconnection is through power or energy as universal quantities across several physical domains, the representation of the pH system offers a cohesive framework for numerical simulations of multi-physical systems \cite{Lohmayer22} and has been applied to model various types of systems, including electrical power systems \cite{GerHRS21,GerSZMS24}, mechanical systems~\cite{Pon24}, fluid dynamics~\cite{Car24} and interacting particle systems~\cite{JacT24}. 

The aim of this paper is to develop a method for identification of nonlinear pH systems based on ML, see also \cite{Cherifi2020} for a recent overview of ML techniques for pH systems.
Various identification methods have been proposed for linear pH systems, i.e.\ when $J$ and $R$ are constant coefficient matrices and $H$ is a quadratic function. For example, there is a pH version of the well-known Dynamic Mode Decomposition (pHDMD) \cite{Morandin2023}, and approaches utilizing linear matrix inequalities (LMIs) \cite{Cherifi2019, GilS18}, optimization \cite{Sch23, Gunther2023, Gunther2024}, subspace identification \cite{Medianu17}, and frequency domain methods using the Loewner framework \cite{Benner2021, Che22, CheB21}.

The extension to pH structure-preserving identification for nonlinear dynamics remains less explored. Recent advances were made in the context of circuits \cite{Najnudel21}, mechanical systems \cite{Desai2021}, using Gaussian Processes \cite{Beckers22,Zaspel2024}, using Autoencoders \cite{Rettberg2024} or in the context of Pseudo-Hamiltonian systems \cite{Eidnes23}. These works have primarily focused on linear pH systems or have only partially addressed nonlinear systems by approximating specific components like the Hamiltonian or dissipation matrices or by using prior assumptions on the structure of the system matrices.  Another direction that is less explored is the identification based on input-output data. Only recently, a method was developed that approximates the input-output behavior of systems while enforcing a port-Hamiltonian structure on a latent dynamics using subspace encoders~\cite{OE-PHNN26}. 

The invariance of pH systems under structure-preserving interconnection was recently used in \cite{Neary2023} to speed up the training process by decomposing a given pH system into pH subsystems which are trained independently and the trained systems are then reassembled. This method was recently extended in \cite{vanOMWTJS24} which approximates an internal port-Hamiltonian system based on input-output data for several time steps.

Current learning-based approaches to nonlinear port-Hamiltonian systems are often tailored to specific architectural choices or fixed structural assumptions, and they provide limited mechanisms for systematically incorporating varying degrees of prior physical knowledge within a single, unified formulation. In this paper, we address this gap by focusing on the identification and learning of nonlinear port-Hamiltonian systems from input-state-output data in one unified framework called ISO-pHNN. Our approach employs multilayer perceptrons (MLPs) designed to maintain the pH structure of the system to approximate each constituent component of the system, namely, the interconnection matrix $J$, the dissipation matrix $R$, the Hamiltonian $H$, and the input matrix $B$. By assigning dedicated MLPs to each matrix and incorporating a parametrization scheme, we ensure that the learned models strictly adhere to the pH structure. This method maintains the system's physical integrity while utilizing neural networks' universal approximation abilities to accurately model intricate nonlinear dynamics. In addition to the first architecture, we also experiment with the newly proposed Kolmogorov-Arnold Networks (KANs) and compare them with MLPs. Compared to the recent approach to identify nonlinear pH systems \cite{vanOMWTJS24}, we use input-state-output data and implement different versions of MLPs and KANs with and without prior knowledge and validate the method for various examples and settings. Our contributions can be summarized as follows:
\begin{itemize}
\item We propose ISO-pHNN, a general flexible framework in which all components of the port-Hamiltonian structure are learned simultaneously.
\item We introduce the concept of physics-informed priors that incorporate physical knowledge to enhance the training of port-Hamiltonian neural networks and can be flexibly used for each individual component.
\item We demonstrate that enforcing port-Hamiltonian structural constraints does not lead to a deterioration of system identification performance. Moreover, using prior information on the structure significantly reduces the error in the learned dynamical system.
\item We explore KAN (Kolmogorov-Arnold Networks) as alternatives to standard MLPs and assess their effectiveness in modeling port-Hamiltonian systems.
\end{itemize}

The paper is structured as follows. In Section~\ref{sec:learning} we present our structure-preserving learning-based approaches using MLPs and KANs. In Section~\ref{sec:training} we describe the training setup including the loss function. Finally, in Section~\ref{sec:examples} the ISO-pHNN framework is applied to various examples that showcase the strengths of the proposed framework.

\section*{Notation}
We write $\mathbb{N}$ for the set of positive integers and $\mathbb{R}$ for the set of real numbers. %Vectors are identified with column vectors. 
For $n,m\in\mathbb{N}$, the space of real $n\times m$ matrices is denoted by $\mathbb{R}^{n\times m}$. For a matrix $M\in\mathbb{R}^{n\times m}$, its transpose is denoted by $M^\top$. If $M\in\mathbb{R}^{n\times n}$ is symmetric, the notation $M\geq 0$ means that $M$ is positive semidefinite. The identity matrix of size $n$ is denoted by $I_n$. 
For a differentiable scalar-valued function $f$, we denote by $\nabla f(x)$ its gradient with respect to the variable $x$. 
The cardinality of a finite set $\mathcal{B}$ is denoted by $|\mathcal{B}|$. Predicted quantities are marked by a hat, for instance $\widehat{\dot{x}}$ and $\widehat{y}$. We use the symbol $\bullet\in\{J,R,H,B\}$ as a placeholder whenever the same statement applies to all four port-Hamiltonian components.

\section{Learning-based identification of port-Hamiltonian systems}
\label{sec:learning}
In this section, we present different variants of ISO-pHNN that combine the general pH framework with neural networks and allow us to identify the nonlinear pH system structure from input-state-output data. More precisely, our aim is to approximate the Hamiltonian $H$ and the system matrices $J$, $R$, and $B$ of the pH system~\eqref{eq:phs} simultaneously by using state-input-output data. To this end, we sample the trajectories of the pH system~\eqref{eq:phs} for various initial values $x_0\in\mathbb{R}^n$ and continuous input signals $u:[0,\infty)\rightarrow\mathbb{R}^m$. To generate a sufficiently rich amount of training data, we sample over different initial conditions $x_0$ and different input trajectories $u$, see Section~\ref{sec:training} for further details.

\subsection{Parametrization of port-Hamiltonian systems}
To preserve the port-Hamiltonian structure of the state-dependent system matrices $J(x)$ and $R(x)$ during the identification process, we build on a~structured representation for the linear case from \cite{Schwerdtner2021,Schwerdtner2022,Sch23} using 
the following reshaping operation that maps vectors to full matrices  
\begin{align*}
&\text{vtf}_{n,m} : \mathbb{R}^{n\cdot m}\rightarrow \mathbb{R}^{n\times m}, & &v=(v_i)_{i=1}^{nm} \mapsto \begin{bmatrix}
v_1 & v_{n+1} & \cdots & v_{m(n-1)+1} \\
v_2 & v_{n+2} & \cdots & v_{m(n-1)+2} \\
\vdots & \vdots &  & \vdots \\
v_n & v_{2n} & \cdots & v_{mn}
\end{bmatrix}.
\end{align*}
This reshaping operation is used in conjunction with 
the state-dependent parameter functions
\begin{align}
\label{def:theta}
\theta_J:\mathcal{X}\rightarrow \mathbb{R}^{n^2},\quad \theta_R:\mathcal{X}\rightarrow \mathbb{R}^{n^2}, \quad \theta_B:\mathcal{X}\rightarrow \mathbb{R}^{mn}, \quad \theta_H:\mathcal{X}\rightarrow\mathbb{R},
\end{align}
to represent the matrix-valued functions $J$, $R$, $B$ and the Hamiltonian $H$ in the pH system representation \eqref{eq:phs} as
\begin{align}
\nonumber
    J(x) &= \text{vtf}_{n,n}(\theta_J(x))^\top - \text{vtf}_{n,n}(\theta_J(x)),\\%\mathcal{M}_{\text{stril}}(\theta_J(x)) - \mathcal{M}_{\text{stril}}(\theta_J(x))^\top, \\
    R(x) &= \frac{1}{\sqrt{n}}\text{vtf}_{n,n}(\theta_R(x)) \, \text{vtf}_{n,n}(\theta_R(x))^\top,\label{eq:direct_parametrization}\\
    B(x) &= \text{vtf}_{n,m}(\theta_B(x)), \nonumber\\
    H(x)&=\theta_H(x), \nonumber 
\end{align}
which, by construction, leads to $J(x) = -J(x)^\top$ and $R(x) = R(x)^\top \geq 0$. 

To improve computational efficiency in the identification of linear pH systems \cite{Schwerdtner2021,Schwerdtner2022}, upper triangular parametrizations are used instead of the full parametrization $\mathrm{vtf}$ in~\eqref{eq:direct_parametrization}. The same strategy could be used to maintain the pH structure for the nonlinear scenario, resulting in a point-wise Cholesky factorization of $R$. However, we do not consider reshaping operations that map to upper triangular matrices as in \cite{Schwerdtner2021,Schwerdtner2022}  here for two reasons. First, we consider approximations of the parameter function $\theta_R$ by deep neural networks where most of the parameters are located in hidden layers, which implies that the parameter reduction using more advanced reshaping operations results in a limited reduction of parameters. Second, for large matrices, the number of summands that appear in the resulting product matrix increases for the bottom rows and rightmost columns, which might introduce scaling problems. Additionally, we add the additional factor $\tfrac{1}{\sqrt{n}}$ into the parametrization  following standard practice in the machine learning community, for example, used in the product of keys and queries in attention layers~\cite{vaswani2023attentionneed}.
For Hamiltonians parametrized by a neural network, one can guarantee that $H(x) \ge 0$ for all $x\in \mathcal{X}$ by applying a positive activation right after the output layer. 
However, in practice we omit  this since we observed that it can destabilize training and instead allow the Hamiltonian to become negative for some points. 
The reason is that we can only guarantee the identified components $J, R, H, B$ to model the true system dynamics in a compact neighbourhood of the training data. In this neighbourhood $H(x) \ge -c$ for some $c > 0$ holds and we can replace $H$ with $H + c$ after training to guarantee that $H$ is positive in the described neighbourhood.
\subsection{Incorporating prior information}
The accuracy of the identification can be further improved by using prior information on the system. In nonlinear system identification it is common to incorporate ansatz functions in the pH system identification. If only limited prior knowledge is available, there are methods such as SINDy \cite{Brunton_2016} where large sets of ansatz functions are prescribed that require sparse identification methods that select only a few of these ansatz functions appearing in the system equations. In contrast to this, we have a very particular pH system structure~\eqref{eq:phs} that cannot be directly incorporated into the SINDy framework.

\begin{figure}[H]
    \centering
    \includegraphics[width=1\textwidth]{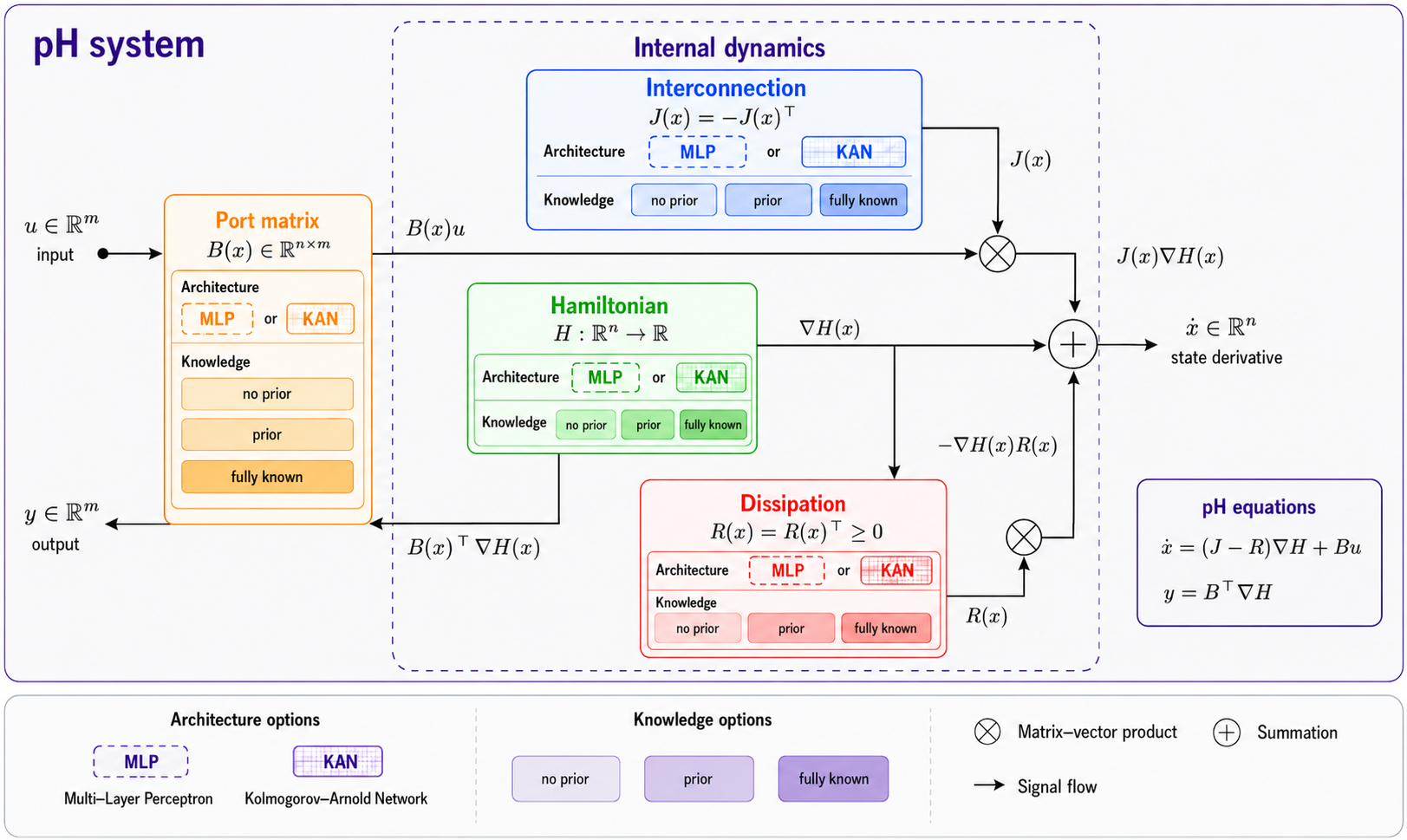}
    \caption{Example of a pH system in terms of its modular components $J$, $R$, $H$, and $B$. Each of these components can be flexibly modeled as an MLP or a KAN; furthermore we can incorporate prior information in the parameter functions.} % Often the interconnection matrix $J$ and the input matrix $B$ are known (green) and the Hamiltonian $H$ and the dissipation $R$ are unknown (blue).}}
    \label{fig:modularity_prior}
\end{figure}

ISO-pHNN enables the integration of prior structural knowledge about the system by allowing a priori specification of some components of the port-Hamiltonian formulation represented by the matrices  $J$, $R$, $H$ and $B$ as illustrated in Figure \ref{fig:modularity_prior}. The method allows us to incorporate partial or complete knowledge of these quantities whenever such information is available. For instance, in many physical systems the interconnection matrix $J$ and the input matrix $B$ are known from first principles or from system design, while the dissipation matrix $R$ and the Hamiltonian $H$ may be uncertain or difficult to model analytically. In such a case, the learning procedure can be restricted to estimating only $R$ and $H$, while $J$ and $B$ remain fixed. This selective parametrization focuses the learning process on the unknown aspects of the dynamics, thereby reducing model complexity, improving data efficiency, and preserving physically meaningful structure in the resulting model. Moreover, the framework allows the user to impose assumptions about the functional dependence of these quantities on the system state. For example, if prior knowledge indicates that the dissipation matrix $R$ is constant, it can be parametrized as a fixed matrix rather than as a state-dependent nonlinear function. Consequently, the model does not spend representational capacity learning unnecessary nonlinear dependencies, which simplifies optimization and further constrains the learned dynamics to stay consistent with known physical properties.

To incorporate prior information in the pH identification we may assume that for some of the pH matrix coefficients   $J$, $R$, $H$ and $B$ there are certain ansatz functions
 \[
 a_{\bullet}(x):=\begin{bmatrix}
      a_{{\bullet},1}(x)\\
    \vdots\\
    a_{{\bullet},N_{\bullet}}(x)
\end{bmatrix},\quad  a_{\bullet,i}: \mathbb{R}^n \rightarrow \mathbb{R}, \quad i=1,\ldots,N_{\bullet},\quad \bullet\in\{J,R,H,B\}.
 \]
Then, as an alternative to \eqref{eq:direct_parametrization} we obtain the following parametrization 
\begin{align}
J(x)&=\text{vtf}_{n,n}(\theta_J^{\rm pri}a_J(x))^\top -\text{vtf}_{n,n}(\theta_J^{\rm pri}a_J(x)),\nonumber  \\ \label{prioir_para}
R(x) &= \frac{1}{\sqrt{n}}\text{vtf}_{ n,n}\left(\theta_R^{\rm pri}a_R(x)\right)^\top\text{vtf}_{n,n}
\left(\theta_R^{\rm pri}a_R(x)\right), \\
H(x)&=\theta_H^{\rm pri}a_H(x),\quad
B(x)= \text{vtf}_{n,m}(\theta_B^{\rm pri}a_B(x)),\nonumber 
 \end{align}
using constant parameter matrices $\theta_J^{\rm pri} \in \mathbb{R}^{n^2\times N_{J}}$, $\theta_R^{\rm pri} \in \mathbb{R}^{n^2\times N_{R}}$, $\theta_H^{\rm pri}\in\mathbb{R}^{1\times N_H}$ and $\theta_B^{\rm pri}\in\mathbb{R}^{nm\times N_B}$ which need to be adjusted to the given input-state-output data. 

For the special case $N_{\bullet}=1$ and $a_{\bullet} = 1$ for all $\bullet\in\{J,R,H,B\}$, we obtain constant $R,J$ and $B$ matrices, which leads to
\begin{align}
    \label{J_R_linear}
J = \text{vtf}_{n,n}(\theta_J^{\rm pri})^\top -\text{vtf}_{n,n}(\theta_J^{\rm pri}),\quad  R = \frac{1}{ \sqrt{n}}\text{vtf}_{n,n}(\theta_R^{\rm pri})^\top \text{vtf}_{n,n}(\theta_R^{\rm pri}).
\end{align}
Moreover, for various pH systems it is reasonable to assume that 
 the Hamiltonian is nonnegative, quadratic and described as
\begin{align}
    \label{quad_Ham}
H(x)=\frac12x^\top Qx,\quad Q= \frac{1}{ \sqrt{n}}\text{vtf}_{n,n}(\theta_Q^{\rm pri})^\top \text{vtf}_{n,n}(\theta_Q^{\rm pri}),
\end{align}
for some parameter $\theta_Q^{\rm pri}\in\mathbb{R}^{n^2}$. The matrix factorization in \eqref{quad_Ham} inherently guarantees that $Q=Q^\top\geq 0$ and therefore no further constraints on $\theta_Q^{\rm pri}$ during the optimization are required \cite{Schwerdtner2021}. If \eqref{J_R_linear} and \eqref{quad_Ham} are assumed, then we obtain the parametrization for linear pH systems from \cite{Schwerdtner2021}, but using a full parametrization instead of upper triangular matrices. 

Within ISO-pHNN, we can flexibly combine parametrizations with and without prior knowledge \eqref{prioir_para} and \eqref{eq:direct_parametrization}, respectively, for each of the different coefficient matrices $J,R,B$ and the Hamiltonian $H$, e.g.\ one is allowed to parametrize $J$ and $B$ via \eqref{eq:direct_parametrization}, $R$ via some ansatz functions \eqref{prioir_para} and the Hamiltonian $H$ as a quadratic function~\eqref{quad_Ham}.

\subsection{Neural network architectures}

The approximation of the parameter functions~\eqref{def:theta} is based on Multi-Layer Perceptrons (MLPs) and Kolmogorov-Arnold Networks (KANs) as illustrated in Figure~\ref{fig:model-architecture}. Each state variable $x_i$ corresponds to a neuron $\mathbf{i}_i$ in the input layer, each component of the parameter function $\theta_{\bullet,i}$ where $\bullet\in\{J,B,R,H\}$ corresponds to a neuron $\mathbf{o}_i$ in the output layer and there are neurons $\mathbf{h}_i^{(j)}$ in the hidden layers. Furthermore, the size of the output layers are given by $n_B=nm$, $n_H=1$, and $n_J=n_R=n^2$.

\begin{figure}[htbp!]
\centering
\begin{tikzpicture}[
    >=Latex,
    line width=0.9pt,
    font=\small,
    neuron/.style={
        circle,
        draw,
        minimum size=8mm,
        inner sep=1pt,
        align=center
    },
    input neuron/.style={
        neuron,
        fill=blue!15,
        draw=blue!60!black
    },
    hidden neuron/.style={
        neuron,
        fill=green!15,
        draw=green!50!black
    },
    output neuron/.style={
        neuron,
        fill=orange!20,
        draw=orange!70!black
    },
    dotsnode/.style={
        draw=none,
        font=\large
    },
    layer label/.style={
        rounded corners=2pt,
        fill=gray!10,
        draw=gray!40,
        inner sep=3pt,
        font=\footnotesize
    },
    conn/.style={
        ->,
        draw=gray!70
    }
]

%------------------------------------------------
% Input variable labels
%------------------------------------------------
\node at (-1.2,5.0) {$x_1$};
\node at (-1.2,3.5) {$\vdots$};
\node at (-1.2,2.0) {$x_n$};

%------------------------------------------------
% Input layer
%------------------------------------------------
\node[input neuron] (i1) at (0,5.0) {$\mathbf{i}_1$};
\node[dotsnode]     (id) at (0,3.5) {$\vdots$};
\node[input neuron] (in) at (0,2.0) {$\mathbf{i}_n$};

%------------------------------------------------
% First hidden layer
%------------------------------------------------
\node[hidden neuron] (h11) at (2.2,5.0) {$\mathbf{h}_1^{(1)}$};
\node[dotsnode]      (h1d) at (2.2,3.5) {$\vdots$};
\node[hidden neuron] (h1n) at (2.2,2.0) {$\mathbf{h}_{n_1}^{(1)}$};

%------------------------------------------------
% Middle hidden layers (symbolic)
%------------------------------------------------
\node[hidden neuron] (hm1) at (4.5,5.0) {$\cdots$};
\node[hidden neuron] (hm2) at (4.5,3.5) {$\cdots$};
\node[hidden neuron] (hm3) at (4.5,2.0) {$\cdots$};

%------------------------------------------------
% Last hidden layer
%------------------------------------------------
\node[hidden neuron] (hd1) at (6.8,5.0) {$\mathbf{h}_1^{(d)}$};
\node[dotsnode]      (hdd) at (6.8,3.5) {$\vdots$};
\node[hidden neuron] (hdn) at (6.8,2.0) {$\mathbf{h}_{n_d}^{(d)}$};

%------------------------------------------------
% Output layer
%------------------------------------------------
\node[output neuron] (o1) at (9.0,5.0) {$\mathbf{o}_1$};
\node[dotsnode]      (od) at (9.0,3.5) {$\vdots$};
\node[output neuron] (on) at (9.0,2.0) {$\mathbf{o}_{n_\bullet}$};

%------------------------------------------------
% Output function labels
%------------------------------------------------
\node at (11.0,5.0) {$\theta_{\bullet,1}(x)$};
\node at (11.0,3.5) {$\vdots$};
\node at (11.0,2.0) {$\theta_{\bullet,n_\bullet}(x)$};

%------------------------------------------------
% Layer captions
%------------------------------------------------
\node[layer label] at (0,0.7)   {Input layer};
\node[layer label] at (4.5,0.7) {Hidden layer(s)};
\node[layer label] at (9.0,0.7) {Output layer};

%------------------------------------------------
% Background boxes
%------------------------------------------------
\begin{scope}[on background layer]
    \node[
        fill=blue!6,
        draw=blue!25,
        rounded corners=6pt,
        fit=(i1)(id)(in),
        inner sep=10pt
    ] {};
    \node[
        fill=green!6,
        draw=green!25,
        rounded corners=6pt,
        fit=(h11)(h1d)(h1n)(hm1)(hm2)(hm3)(hd1)(hdd)(hdn),
        inner sep=10pt
    ] {};
    \node[
        fill=orange!8,
        draw=orange!30,
        rounded corners=6pt,
        fit=(o1)(od)(on),
        inner sep=10pt
    ] {};
\end{scope}

%------------------------------------------------
% Connections: input -> first hidden
%------------------------------------------------
\draw[conn] (i1) -- (h11);
\draw[conn] (i1) -- (h1d);
\draw[conn] (i1) -- (h1n);

\draw[conn] (in) -- (h11);
\draw[conn] (in) -- (h1d);
\draw[conn] (in) -- (h1n);

%------------------------------------------------
% Connections: first hidden -> middle symbolic hidden
%------------------------------------------------
\draw[conn] (h11) -- (hm1);
\draw[conn] (h11) -- (hm2);
\draw[conn] (h11) -- (hm3);

\draw[conn] (h1n) -- (hm1);
\draw[conn] (h1n) -- (hm2);
\draw[conn] (h1n) -- (hm3);

%------------------------------------------------
% Connections: middle symbolic hidden -> last hidden
%------------------------------------------------
\draw[conn] (hm1) -- (hd1);
\draw[conn] (hm1) -- (hdd);
\draw[conn] (hm1) -- (hdn);

\draw[conn] (hm3) -- (hd1);
\draw[conn] (hm3) -- (hdd);
\draw[conn] (hm3) -- (hdn);

%------------------------------------------------
% Connections: last hidden -> output
%------------------------------------------------
\draw[conn] (hd1) -- (o1);
\draw[conn] (hd1) -- (on);

\draw[conn] (hdn) -- (o1);
\draw[conn] (hdn) -- (on);

\end{tikzpicture}
\caption{Generic MLP or KAN architecture used for approximation of the parameter functions $\theta_{\bullet}$ where $\bullet\in\{J,R,H,B\}$.}
\label{fig:model-architecture}
\end{figure}

MLPs are the most basic type of deep neural network. They are fully connected feedforward artificial neural networks~\cite{Hastie2009}, but despite their simplicity, they inherit a universal approximation property that makes them capable of approximating any continuous function in a compact domain uniformly with any prescribed accuracy, provided that a sufficient amount of data is available, see~\cite{Cybenko1989,Hornik1989}. 

MLPs can be mathematically defined as follows
\begin{equation}
    \varphi_l = \sigma_l\left(W_l\varphi_{l-1} + b_l\right),\quad l=1,\ldots,d+1.
\end{equation}
Here, $l$ denotes the index of the current layer, $n_l$ is the number of neurons in the $l$-th layer and $\varphi_{l}\in\dR^{n_l}$. The matrix $W_{l}\in\dR^{n_l\times n_{l-1}}$ represents the weight parameters of this layer, $b_{l}\in\dR^{n_{l}}$ is the associated bias vector, and $\sigma_l:\dR^{n_{l-1}}\rightarrow\dR^{n_l}$ is a (typically nonlinear) activation function used to capture the inherent nonlinear behaviour of the system (when present). Moreover, the layers $l=0$ and $l=d+1$ are called input and output layer, respectively. 

When fully assembled, the MLP transforms an input feature vector $\varphi_0 \in \mathbb{R}^{n_0} $ into an output vector $\varphi_{d+1} \in \mathbb{R}^{n_{d+1}}$ which is defined recursively as 
\[
\text{MLP}_{\phi}: \mathbb{R}^{n_0} \rightarrow \mathbb{R}^{n_{d+1}}, \quad \varphi_0 \mapsto \varphi_{d+1}=\sigma_{d+1}(W_{d+1}\varphi_{d} + b_{d+1}),
\]
starting with $\varphi_1 = \sigma_1\left(W_1\varphi_{0} + b_1\right)$ and 
where $\phi =\{(W_1,b_1),\ldots,(W_{d+1},b_{d+1})\}$ represents the set of trainable parameters.

In the ISO-pHNN architecture, we train the parameters of several MLPs to approximate the parameter functions
\[
\theta_\bullet(x)\approx\text{MLP}_{\bullet,\phi}(x),\quad \bullet\in\{J,R,H,B\}.
\]
Although separate neural networks are used for the different components, they are trained jointly through a single loss function. Each neural network is then used to recover the system matrices and then evaluate the nonlinear pH model directly.

In addition to MLPs, we also explore multilayer KANs \cite{Liu2024}. These architectures are inspired by the Kolmogorov--Arnold representation theorem. In the formulation recalled in \cite{Liu2024}, for a sufficiently regular function $f:[0,1]^n\to\mathbb{R}$, KAN functions are of the form
\begin{equation}
\label{KAN_representation}
f(z_1,\ldots,z_n)=\sum_{q=1}^{2n+1}\Phi_q\!\left(\sum_{p=1}^n \phi_{q,p}(z_p)\right),
\end{equation}
where $\phi_{q,p}:[0,1]\to\mathbb{R}$ and $\Phi_q:\mathbb{R}\to\mathbb{R}$ are univariate functions. Thus, the theorem is stated on the domain $[0,1]^n$ and motivates the use of univariate learnable functions to represent multivariate mappings.
This theorem should be distinguished from the practical KAN architecture and its implementation. In KANs, learnable univariate functions are placed on the edges of the network, and deeper KANs are obtained by stacking layers of such functions. Hence, the learnable edge functions used in practice play the role of the univariate functions appearing in the Kolmogorov--Arnold representation, although the resulting deep architecture is more general than the original two-layer theorem-based representation.

In the implementation of \cite{Liu2024}, each learnable univariate edge function is parametrized as
\begin{equation}
\label{kan_activation}
\Psi(z)=w_b\,b_{\mathrm{act}}(z)+w_s\,\mathrm{spline}(z), \qquad
\mathrm{spline}(z)=\sum_i c_i N_i(z),
\end{equation}
where $b_{\mathrm{act}}$ is a fixed basis activation function, $N_i$ are $B$-spline basis functions, and $c_i$, $w_b$, and $w_s$ are trainable parameters. Therefore, $\Psi$ is the parametric form used in the implementation to represent each learnable univariate edge function.

The spline part is defined on a bounded grid, i.e., a partition of the input interval into subintervals on which the $B$-spline basis functions are constructed. The grid size denotes the number of such subintervals. In particular, this grid-based implementation is not tied to the theorem domain $[0,1]$. By reparametrizing the functions in \eqref{KAN_representation}, we can choose the spline grid to be an arbitrary compact interval. In our implementation we use the grid range $[-1,1]$. In all our experiments, we use cubic splines and a grid size of $4$. Consequently, all $B$-spline basis functions have the same degree, and the number of basis functions is determined by the spline degree and the grid size.

Learning functions instead of individual weights in MLPs introduces significant computational challenges. Nonetheless, \cite{Liu2024} demonstrates that for some applications only a small number of neurons in KANs are required to match the precision of a larger MLP. Fewer neurons may also improve network interpretability. In contrast to standard MLPs, KANs learn adaptive univariate functions along edges, which can make learned nonlinear relations easier to inspect and visualize. Recent work has further shown that such representations may support sparsification and symbolic post-processing, thereby providing more interpretable descriptions of the learned dynamics \cite{Koenig2024KANODE}. A similar interpretability study of pH systems would be interesting and will be pursued in future work.

\section{Training setup}
\label{sec:training}
In this section, we discuss the data used to train the neural networks, the baseline, and the loss function. We also discuss the error measures used to compare the different models. All experiments were conducted in PyTorch~\cite{paszke2019pytorchimperativestylehighperformance}. Following the concept from \cite{vanOMWTJS24}, we utilize PyTorch's autograd function to compute the gradient of $\theta_H$ at input $x \in \mathbb{R}^n$, which approximates $\nabla H(x)$. 

\subsection{Input-state-output data generation}
\label{sec:data_generation}
The generation of data is similar to \cite{Neary2023}. We uniformly sample initial states $x_{i,0} \sim U(x_{i, \mathrm{min}}, x_{i, \mathrm{max}})$, $i=1,\ldots,n$. The trajectories are generated by using SciPy's \texttt{odeint} function  \cite{virtanen2020scipy}.
The choice of input signal determines a range in which the states are typically located. It is thus important to choose the $x_{i, \mathrm{min}}, x_{i, \mathrm{max}}$ within this range to avoid outliers right at the beginning of the trajectories.
Besides that, the input signal has a much higher significance than the initial condition since the input signal frequently excites the system.
We train several variants of ISO-pHNN for validation. To show that our method performs well for various data sizes, we consider $N_{\mathrm{tra}}=4,8,16,32,64$ as the number of training trajectories for each of the example systems which are introduced later. Moreover, we use $N_{\mathrm{tra,v}} = 100$ fixed validation trajectories. In order to harden the task of long-time forecasting, each training trajectory represents $10\mathrm{s}$ while each validation trajectory represents $100\mathrm{s}$. We use a sampling interval of $\Delta t = 0.01\, \mathrm{s}$ in each case, resulting in $1000$ data points per training trajectory and $10000$ data points per validation trajectory. Note that the validation dataset is much larger than the training dataset. Since we use simulation data, this has no disadvantage and allows us to get reliable validation results in all considered experiments. For the validation data, we use different random conditions and different randomly sampled input signals.

We follow \cite{vanOMWTJS24} to sample the input signal
    \begin{align}
    \label{def:input}
    u_i(t) =
    \sum_{k=1}^N
    \underbrace{a \sin(2\pi k f_{0} t + \phi^k_i)}_{u^k_i(t)}, \quad i = 1, \dots, m
    \end{align}
    with $f_0 = 0.1$ Hz, $N=40$, amplitude $a$, and random phases $\phi_i^k$ that are uniformly sampled from $[0,2\pi)$. The index $k$ represents the $k$-th term in the summation over the sinusoids. The training targets are the time derivatives of the states $\dot{x}$ and the outputs $y$, i.e., the quantities that the identified model is trained to reproduce. Using the equations of the pH system \eqref{eq:phs}, this also allows us to determine the values of the derivatives $\dot x$. However, in practical situations, it is common to estimate $\dot{x}$ from observed trajectories. A~significant portion of the SINDy project \cite{Brunton_2016} has been devoted to this. For simplicity, we used instead the values given by the system equation with true coefficients.
    
    \subsection{Loss function and optimization}
The dataset for training consists of trajectories $\{x^k(t_j)=x(t_j;x_0^k,u^k) ~|~ j=1,\ldots,N_t\}$, generated from different initial conditions $x_0^k$ and input signals $u^k = (u^k(t_1), \dots, u^k(t_{N_t}))$, $k=1,\ldots,N_{\rm tra}$, as outlined in Section~\ref{sec:data_generation}. %Furthermore, we abbreviate $x(\cdot;x_{0,k},u_k)$ by $x_k(\cdot)$.
All trajectories in the dataset are concatenated 
\begin{align}
\label{concatenated}
(x^1(t_1),\ldots, x^1(t_{N_t}),\ldots,x^{N_{\rm tra}}(t_1),\ldots ,x^{N_{\rm tra}}( t_{N_t}))\in\dR^{N_tN_{\rm tra}n}
\end{align}
%where $i$ selects the trajectory for a given initial state and input pair $(x_{i,0},u_i)$, $j=1,\ldots n$ represents the state and $k=0,\ldots,N$ .
and from these concatenated trajectories we form the dataset. In order to lessen the computational demands of weight optimization, we employ a common deep learning strategy that creates a minibatch $\mathcal{B}$ out of the entire data set~\eqref{concatenated}  that is used in each training iteration. This is done by randomly selecting $|\mathcal{B}|>0$ entries of \eqref{concatenated}
 \[
%\tau_{\mathcal{B}}: 
b\mapsto (j(b),k(b)),\quad b=1,\ldots,|\mathcal{B}|
 \]
where $j(b) \sim \mathcal{U}(\{1, \dots, N_t\}), k(b) \sim\mathcal{U}(\{1, \dots, N_{\rm tra}\})$ are sampled uniformly.
Moreover, using the system equations~\eqref{eq:phs}, the data set uniquely specifies the values of the state-derivatives $\dot{x}^k(t_j)=\dot{x}(t_j;x_0^k,u^k)$ and the corresponding outputs $y^k(t_j)=y(t_j;x_0^k,u^k)$. By default, we min-max scale states before processing them with our neural networks.

More specifically, given $1 \le i \le n$, $1 \le q \le m$, we compute  
    \begin{align*}
    M_{x_i} = \max\limits_{\substack{
    k=1,\dots,N_{\mathrm{tra}} \\
    j = 1, \dots, N_t
    }} x_{i}^k(t_j),\qquad 
    m_{x_i} = \min\limits_{\substack{
    k=1,\dots,N_{\mathrm{tra}} \\
    j = 1, \dots, N_t
    }} x_{i}^k(t_j),
\end{align*}
and then determine
\begin{align*}&(\tilde{x}^{k(b)}_1(t_{j(b)}), \dots, \tilde{x}^{k(b)}_n(t_{j(b)})) \\&\quad= \biggl(\frac{2(x_1^{k(b)}(t_{j(b)}) - m_{x_1})}{M_{x_1} - m_{x_1}} - 1,\dots, \frac{2(x_n^{k(b)}(t_{j(b)}) - m_{x_n})}{M_{x_n} - m_{x_n}} - 1\biggr),
%\\ (\tilde{u}_1, \dots, \tilde{u}_m) &= (\frac{u_1 - \tilde \mu_{u_1}}{\tilde \sigma_{u_1}},\dots, \frac{u_m - \tilde \mu_{u_m}}{\tilde \sigma_{u_m}}).
\end{align*}
and use $(\tilde{x}^{k(b)}(t_{j(b)}), u^{k(b)}(t_{j(b)}))$ as input to the neural networks.

The reason why we do not normalize $y, \dot{x}, u$ before processing these quantities is that the normalization of these quantities may alter the system class considered. For example, suppose that $\tilde{u} \in \mathbb{R}^m$ is computed from $u$ via normalizing (either min-max scaling or standard scaling), then there exist an invertible matrix 
$A \in \mathbb{R}^{m\times m}$ and a vector $b\in \mathbb{R}^m$ 
such that $\tilde{u} = A u + b$. Hence, \eqref{eq:phs} with $\tilde{u}$ as the input instead of $u$ would read
\begin{align*}
    \dot{x}(t) &= \left(J(x(t)) - R(x(t))\right) \nabla H(x(t)) + B(x(t))A^{-1}\tilde{u}(t) - B(x(t))A^{-1}b \\
    y(t) &= B(x(t))^\top \nabla H(x(t)), \quad t\geq 0,% + \left(N(x) + S(x)\right)u.
\end{align*}
This system does not need to be port-Hamiltonian, in fact not even passive as one can see by considering simple examples.

Similar problems can occur if $\dot{x}$ or $y$ are normalized which is the reason why we generally only normalize $x$. Normalizing $x$, however, is not expected to cause such problems since the neural networks can invert this normalization if necessary. However, when a prior is being used, this is not the case any more, thus we do not normalize $x$ which means that we set $(\tilde{x}_1^{k(b)}(t_{j(b)}), \dots, \tilde{x}_n^{k(b)}(t_{j(b)})) = ({x}_1^{k(b)}(t_{j(b)}), \dots, x_n^{k(b)}(t_{j(b)}))$ in that case.

The minibatch is then given by \[( \tilde x^{k(b)}(t_{j(b)}),\dot x^{k(b)}(t_{j(b)}), u^{k(b)}(t_{j(b)}), y^{k(b)}(t_{j(b)}))_{b=1}^{|\mathcal{B}|}.\]

The corresponding predictions for $\dot{x}$ and $y$ of our model are indicated by $\widehat{\dot{x}}$ and $\widehat{y}$, respectively. 

Our objective is to minimize the error that is normalized over the training minibatch for each target, i.e. 
    {
    \begin{align}
        \mathcal{L}(\hat{\dot{x}}, \dot{x}, \hat{y}, y) &=         \frac{1}{|\mathcal{B}|}\sum\limits_{b = 1}^{|\mathcal{B}|} 
        \sum_{i=1}^n \frac{|\dot{x}^{k(b)}_i(t_{j(b)}) - \widehat{\dot{x}}^{k(b)}_i(t_{j(b)})|^2}{\tilde{\sigma}_{\dot{x}_i}^2} \nonumber\\
        &\qquad+ \gamma \frac{1}{|\mathcal{B}|}\sum\limits_{b = 1}^{|\mathcal{B}|} \sum_{i=1}^m \frac{|y_{i}^{k(b)}(t_{j(b)}) - \widehat{y}_i^{k(b)}(t_{j(b)})|^2}{\tilde{\sigma}_{y_i}^2} %\text{\large )}
        \label{eq:Beta}
    \end{align}
    }
    with $\tilde \sigma_{\dot{x}_i}$ and $\tilde \sigma_{y_j}$ representing the standard deviation of $\dot{x}_i$ and $y_i$ over the current training minibatch, respectively, i.e. \begin{align*}
        \tilde \sigma_{\dot{x}_i} = \sqrt{\frac{1}{|\mathcal{B}|-1} \sum\limits_{b=1}^{|\mathcal{B}|}\left( \dot{x}_{i}^{k(b)}(t_{j(b)}) - \frac{1}{|\mathcal{B}|}\sum\limits_{b=1}^{|\mathcal{B}|} \dot{x}_{i}^{k(b)}(t_{j(b)}) \right)^2 }\\
        \tilde \sigma_{y_i} = \sqrt{\frac{1}{|\mathcal{B}|-1} \sum\limits_{b=1}^{|\mathcal{B}|}\left( y_{i}^{k(b)}(t_{j(b)}) - \frac{1}{|\mathcal{B}|}\sum\limits_{b=1}^{|\mathcal{B}|} y_{i}^{k(b)}(t_{j(b)}) \right)^2 }
    \end{align*} 
    where $\gamma > 0,$ and \[( \tilde x^{k(b)}(t_{j(b)}),\dot x^{k(b)}(t_{j(b)}), u^{k(b)}(t_{j(b)}), y^{k(b)}(t_{j(b)}))_{b=1}^{|\mathcal{B}|}\] is the current training minibatch.
    Here $\gamma>0$ weights the output loss and we typically use $\gamma=0.25$ because correctly predicting the state derivatives is more crucial for long-term forecasting.
    In our experiments, we noticed that the validation error does not strongly depend on the exact value of $\gamma$. All the choices of $\gamma$ considered resulted in the same order of magnitude in the error measures considered. 
    
    Commonly, different states have different orders of magnitude and, consequently, all states have to be rescaled to have the same order of magnitude, e.g., as in \cite{Rettberg2024}. For the considered examples, we found that our setup works well without rescaling any relevant quantities by simply modifying the loss function to account for the standard deviation of each target as described above.
    Now define 
    \[\sigma_{\dot{x}_i} = \sqrt{\frac{1}{N_{\mathrm{tra, v}}\cdot N_v-1} \sum\limits_{j=1}^{N_v}\sum\limits_{k=N_{\mathrm{tra}}+1}^{N_{\mathrm{tra}}
    +N_{\mathrm{tra, v}}}\left( \dot{x}^k_{i}(t_j) - \frac{1}{N_{\mathrm{tra, v}}\cdot N_v}\sum\limits_{j=1}^{N_v}\sum\limits_{k=N_{\mathrm{tra}}+1}^{N_{\mathrm{tra}} + N_{\mathrm{tra, v}}} \dot{x}^k_{ i}(t_j) \right)^2 }\]
    where $\{x^k(t_j)=x(t_j;x_{0}^k, u^k) ~|~ j=1,\ldots,N_v\}$ for $k=N_{\mathrm{tra}}+1, \dots, N_{\mathrm{tra}} + N_{\mathrm{tra, v}}$ are the validation trajectories with $\{\dot{x}^k(t_j)=\dot{x}(t_j;x_{0}^k,u^k) ~|~ j=1,\ldots,N_v\}$ being the corresponding time-derivatives.
    
    Our main evaluation objective is the following normalized mean absolute error (NMAE)  
    \[
    \mathrm{NMAE}(\dot{x}, \widehat{\dot{x}}) = \frac{1}{n \cdot N_{\mathrm{tra, v}}\cdot N_v}\sum_{i=1}^n\sum\limits_{j=1}^{N_v}\sum\limits_{k=N_{\mathrm{tra}}+1}^{N_{\mathrm{tra}}+N_{\mathrm{tra, v}}} \frac{|\dot{x}_{i}^k(t_j) - \widehat{\dot{x}^k}_{ i}(t_j)|}{\sigma_{\dot{x}_i}}.
    \] We use here a normalized MAE, because we do not rescale the state derivatives $\dot{x}$, which implies that derivatives of different states can have vastly different magnitudes.
    
    We point out that the error on the derivatives of the states is more important for long-term forecasting compared to the error on the output, as these accumulate during forecasting. 
    
    Incorporation of the system output is not only being used as an auxiliary target to improve the prediction of $\dot{x}$, but also because of the system-theoretic properties related to system interconnection, which require access to correct system outputs.

For the activation function, we use $\swish:\mathbb{R}\rightarrow\mathbb{R}$, see \cite{elfwing2018sigmoid}
, that is given by 
\begin{align}
    \label{def:swish}
\swish(z):=\frac{z}{1+e^{-z}}.
\end{align}
We use this activation function instead of the basic $\mathrm{ReLU}(z) = \max(0, z)$ function, because the neural network for the Hamiltonian needs to have two derivatives to allow training using backpropagation which would not be the case using $\mathrm{ReLU}$.
Moreover, we use layer normalization~\cite{lei2016layer} at hidden layers. 
The efficient open-source code from \cite{EffKAN24} is used to implement KANs in PyTorch.

    To optimize the parameters, we use the Adam optimizer \cite{kingma2014adam} with decoupled weight decay \cite{Loshchilov2017}, commonly referred to as AdamW. We employ cosine annealing \cite{loshchilov2017sgdrstochasticgradientdescent} with warmup \cite{goyal2018accuratelargeminibatchsgd} to schedule the learning rate.

\subsection{Forecasting with the identified models}
To verify that the proposed method is good at long-time forecasting, we forecast the validation trajectories with the given initial conditions and the input signals using the identified system equations. To do this, we use the RK5(4) scheme originally proposed in \cite{dormand1980family}. We find that this scheme works well for our purposes due to its high order of accuracy. A disadvantage is, however, that it does not preserve the structure of the considered class of port-Hamiltonian systems. Experiments using the implicit midpoint rule on the other hand, showed a significant deterioration of the results due to the lower order of accuracy of this scheme. We refer to the recent work \cite{linares2026controlled} for a detailed discussion of the integration scheme used for port-Hamiltonian system identification.
\subsection{Baseline}
We compare the identified pH systems with a non-physics-informed baseline. This baseline can be seen as a black-box system model defined as \begin{align*}
        \begin{bmatrix}\dot{x}_{\mathrm{pred}}\\ y\end{bmatrix} = f_{\mathrm{base}}(x, u).
    \end{align*}
    The function $f_{\mathrm{base}}$ is parametrized using an MLP with the same depth as those used in the pH structure-preserving method. To ensure a fair comparison, we scaled the width of the MLPs so that the total parameter count equals that of our default strategy. 
\section{Numerical experiments}
\label{sec:examples}

In this section, we demonstrate ISO-pHNN for various types of port-Hamiltonian systems, each showcasing different aspects of system nonlinearity. Specifically, we examine examples that are nonlinear either in the skew-symmetric matrix $J$ or the dissipation matrix $R$, and may feature quadratic or non-quadratic Hamiltonians~$H$. A computer equipped with an AMD Ryzen 9 5900X 12-Core processor and 32 GB of RAM was used to conduct 
most of the experiments. 

For example, on this computer, training KANs for $1000$ epochs or MLPs for $4000$ epochs and using $32$ trajectories takes approximately one hour, independently of the example systems we considered.
 A validation dataset consisting of $100$ trajectories each representing a $100$ seconds long simulation of the system
 was generated.  Generating the whole dataset took up to $2$~minutes depending on the  example system. Using a model to predict these trajectories takes approximately $5$~minutes.
We also used a high performance cluster with 268 worker nodes,
17152 Xeon Gold 6238R CPU cores and 931 TB of parallel storage for hyperparameter tuning on each example system.

\subsection{Default hyperparameters}
In all  the experiments, we use the batch size $|\mathcal{B}|=256$, learning rate $0.001$, betas $(0.9, 0.999)$ and $\varepsilon = 10^{-8}$  as defined in \cite{kingma2014adam}.
By default, we use the weight decay $0.01$ which is set to $0$ for components where prior is used to stabilize training. Throughout the experiments, we never normalize $u, \dot{x}, y$ and only normalize $x$ using min-max scaling if no prior is being used. By default, $3$ hidden layers with $16$ neurons each are being used for MLP and $3$ hidden layers with a dimension of $4$ for KAN. For the remaining hyperparameters of the used KAN implementation, we always use the grid size $4$, spline order $3$, noise scale $0.1$, base scale $1$, spline scale $1$, base activation SiLU, grid epsilon $0.02$ and grid range $[-1, 1]$.
By default, we train for $4000$ epochs when MLP is used and $1000$ when KAN is used, since one epoch takes longer for KAN. Assuming that $16$ trajectories are used, this amounts to approximately $\frac{4000 \cdot 16 \cdot 1000}{256} = 250000$ training iterations  for MLPs. We find that this high number of iterations is necessary to achieve close to optimal results. 
We typically use the output weight $\gamma = 0.25$. 
A detailed study of the dependency and scaling behavior of the accuracy of the identified models in dependence with the resources, such as the amount of training data, the number of neurons in the MLPs, and computational effort, is conducted in~\cite{roschkowski2026neuralscalinglawslearningbased}.

\subsection{Mass-spring-damper system}

We consider a mass-spring-damper system from \cite{Neary2023} which is composed of a mass $m_1$ that is attached to a wall and a second mass $m_2$ by two springs with corresponding spring constants $k_1,k_2>0$. The relative elongation $q_i$ and the momentum $p_i$ of the masses lead to the states $(x_1,x_2)=(q_1,p_1)$ and $(x_3,x_4)=(q_2,p_2)$, respectively. A detailed explanation along with an illustrative figure can be found in \cite{Neary2023}.
Furthermore, as proposed in \cite{Lopes2015}, we introduce a nonlinear damping force $F_i=b_im_i^{-3} p_i^3$, $i=1,2$, for some $b_i\geq 0$ acting on each of the subsystems. The Hamiltonian for each 
subsystem $H_i(x_i)$ is given by
\[
H_i(x_i) = \frac{p_i^2}{2m_i} + \frac{k_i q_i^2}{2},
\]
which leads to the Hamiltonian $H=H_1+H_2$ for the interconnected system which is given by 
\[
\dot{x} = (J - R(x)) \nabla H(x) + B u(t),
\]
where the skew-symmetric matrix $J$, the damping matrix $R$, and the input matrix $B$ are given by 
\[
J = \begin{bmatrix}
0 & 1 & 0 & 0 \\
-1 & 0 & 1 & 0 \\
0 & -1 & 0 & 1 \\
0 & 0 & -1 & 0
\end{bmatrix}, \quad R(x) = \begin{bmatrix}
0 & 0 & 0 & 0 \\
0 & \frac{b_1 p_1^2}{m_1^2} & 0 & 0 \\
0 & 0 & 0 & 0 \\
0 & 0 & 0 & \frac{b_2 p_2^2}{m_2^2}
\end{bmatrix}, \quad B = \begin{bmatrix}
0 & 0 \\
1 & 0 \\
0 & 0 \\
0 & 1
\end{bmatrix}
\]
and $u$ represents forces that are acting on each of the masses. The amplitude of the input is set to $a = 0.4$ in order to rescale the combined sinusoidal signals. Moreover, we sample the initial states uniformly from $\begin{pmatrix}
q_{1,0}, p_{1,0}, q_{2,0}, p_{2,0}
\end{pmatrix} \sim [-0.5, 0.5]^4$.

To obtain the simulation data for the learning-based system identification task, we choose the model parameters that are listed in Table~\ref{tab:mass_spring}. 
\begin{table}[h!]
\centering
\begin{tabular}{|c |c |c |c |}
%\toprule
\hline
symbol& meaning & value & unit \\
\hline\hline
$m_1$ & mass 1 & 1 & $[{\rm kg} ]$ \\ $m_2$ & mass 2 & 1.5 & $[{\rm kg} ]$ \\ 
$b_1,b_2$ & damping coefficient & 2 & $[\rm Ns^3/m^3]$ \\
$k_1$ & spring constant 1 & 1 & $[{\rm N/m}]$ \\
$k_2$ & spring constant 2 & 0.1 & $[ {\rm N/m}]$ \\
\hline
%\bottomrule
\end{tabular}
\caption{System parameters for the interconnected mass spring system}
\label{tab:mass_spring}
\end{table}
In the above example, we considered a quadratic Hamiltonian. It is possible to modify the example to obtain non-quadratic Hamiltonians, by applying hardening constraints to the spring, as discussed in \cite{Lopes2015}, or by considering a Duffing-type oscillator, as shown in \cite{Dan24}.

\begin{figure}[H]
    \centering
    \includegraphics[width=1\textwidth]{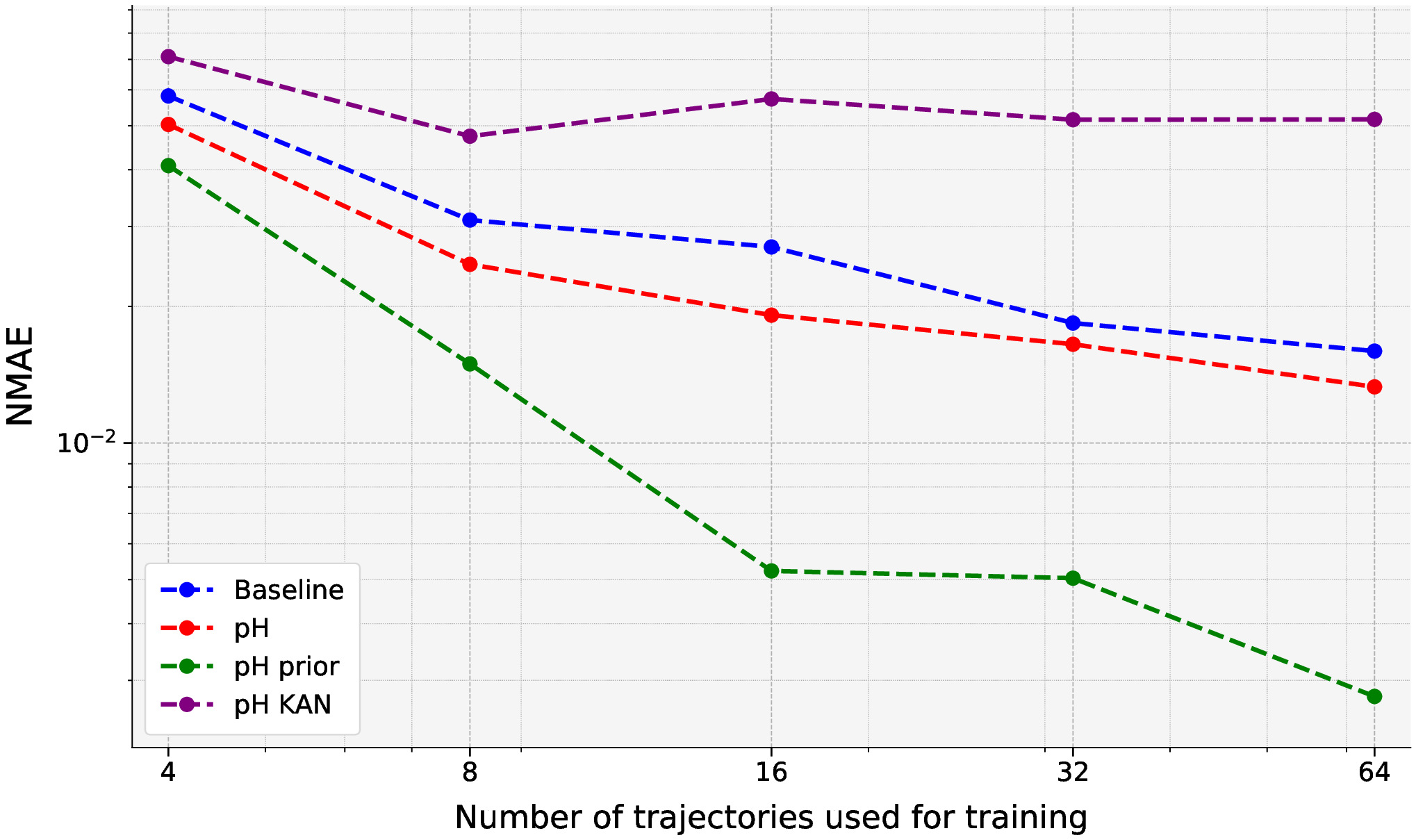}
    \caption{Average validation error for the mass-spring-damper example as a function of varying quantities of trajectories employed in training a baseline model, a default pH setup, the KAN implementation and a pH setup incorporating prior knowledge.}
    \label{fig:Scaling_Spring}
\end{figure}

As a first experiment, we consider how the proposed method scales with more data. Multiple experiments were carried out for the training trajectory values $4, 8, 16, 32, 64$. We try $500, 1000, 2000, 4000, 8000$ epochs and $2, 4, 6, 8, \dots, 26$ for the model hidden dimension to train the models. This hidden dimension is the total number of neurons in the hidden layers of our neural networks. For each trajectory count we report the score of the best training configuration, distinguished by the model hidden dimension and number of epochs used. This allows us to test our method under different data constraints. The experimental conditions are summarized in Table~\ref{tab:params_scaling_spring}. In this case, no noise was added to the trajectories. Therefore, the signal-to-noise ratio (SNR), which quantifies the amount of noise present in the trajectories, is reported as `na'.
The resulting normalized MAE is shown in Figure~\ref{fig:Scaling_Spring}. The error is averaged over the $N_{\mathrm{tra,v}} = 100$ different test trajectories.
The validation error tends to decrease as the number of trajectories used for training increases. The pH implementation is compared with a baseline without any prior physical information. In this experiment, the pH method slightly outperforms the baseline. In addition, we also train a pH prior that incorporates prior information about the model by assuming that the matrix $J$ in the pH model is constant, $B$ is constant and the Hamiltonian to be quadratic. In this case we do not use weight decay in the parameters of components with prior knowledge. The resulting pH prior model achieves a lower error than both the general pH model and the baseline. This confirms that  incorporating physical prior information into ISO-pHNN can further reduce the error.
We also compare the MLP-based architecture with a KAN-based approach. As shown in Figure~\ref{fig:Scaling_Spring}, the KAN-based approach underperforms in this setting. A possible explanation is the different spectral bias of KANs: recent work suggests that KANs are less biased toward low frequencies than MLPs and may be better suited to learning higher-frequency components, which may be less beneficial in the present regime \cite{WanSLH25}. As we will see later, however, in higher-frequency settings such as the PMSM benchmark, KANs can instead outperform MLPs.

\begin{table}[h!]
\centering
\resizebox{\textwidth}{!}{%
\begin{tabular}{|c |c |c |c |c |c |c |c |c |c| c| }
%\toprule
\hline
arch. & $N_{\mathrm{tra}}$& ep. & $\gamma$ &depth &dim & $J$ & $R$ & $H$ & $B$ & SNR\\
\hline\hline
pH & $4,8,16,32,64$& $500, 1000, 2000, 4000, 8000$ & $0.25$  &$3$&$4, 6, \dots, 26$ & MLP & MLP & MLP & MLP &na\\
\hline
pH KAN & $4,8,16,32, 64$& $500, 1000, 2000, 4000, 8000$ & $0.25$  &$3$&$2, 3, 4, 5, 6, 7$ & KAN & KAN & KAN & KAN& na\\
\hline
pH prior & $4,8,16,32,64$& $500, 1000, 2000, 4000, 8000$ & $0.25$ &$3$ &$4, 6, \dots, 26$ & const. & MLP & quadr. & const. & na\\
\hline
baseline & $4,8,16,32,64$& $500, 1000, 2000, 4000, 8000$ & $0.25$ &$3$ &$4, 6, \dots, 26$ & na & na & na & na &na\\
\hline
%\bottomrule
\end{tabular}
}
\caption{Training parameters used in Figure~\ref{fig:Scaling_Spring}.}
\label{tab:params_scaling_spring}
\end{table}

\begin{figure}[H]
    \centering
    % First Subfigure (Left)
    \begin{subfigure}[b]{0.49\textwidth}
        \centering
        \includegraphics[width=\textwidth]{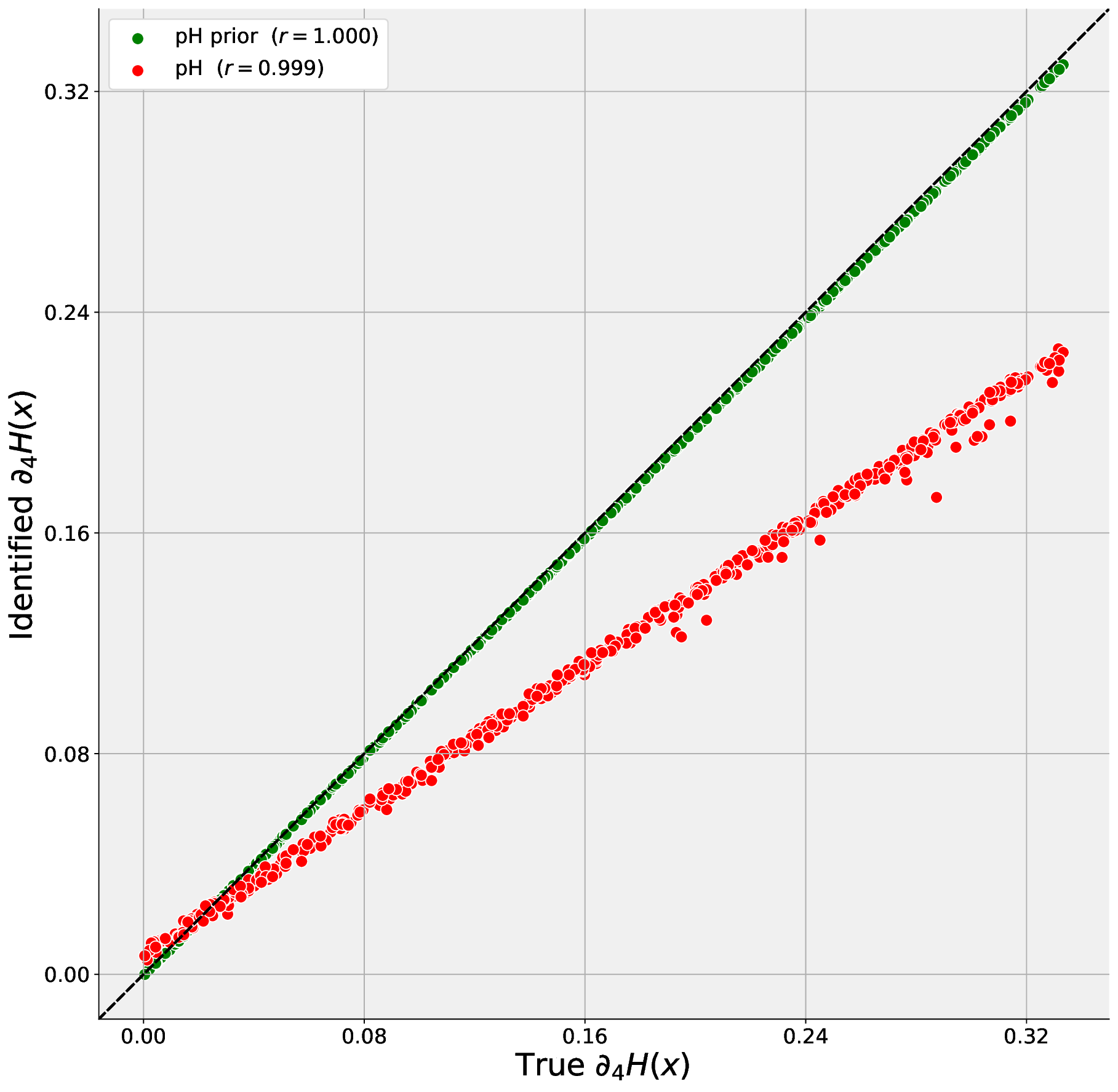}
        \caption{The derivative of $H$ with respect to $x_4$}
        \label{fig:Compare_H}
    \end{subfigure}
    \hfill
    % Second Subfigure (Right)
    \begin{subfigure}[b]{0.49\textwidth}
        \centering
        \includegraphics[width=\textwidth]{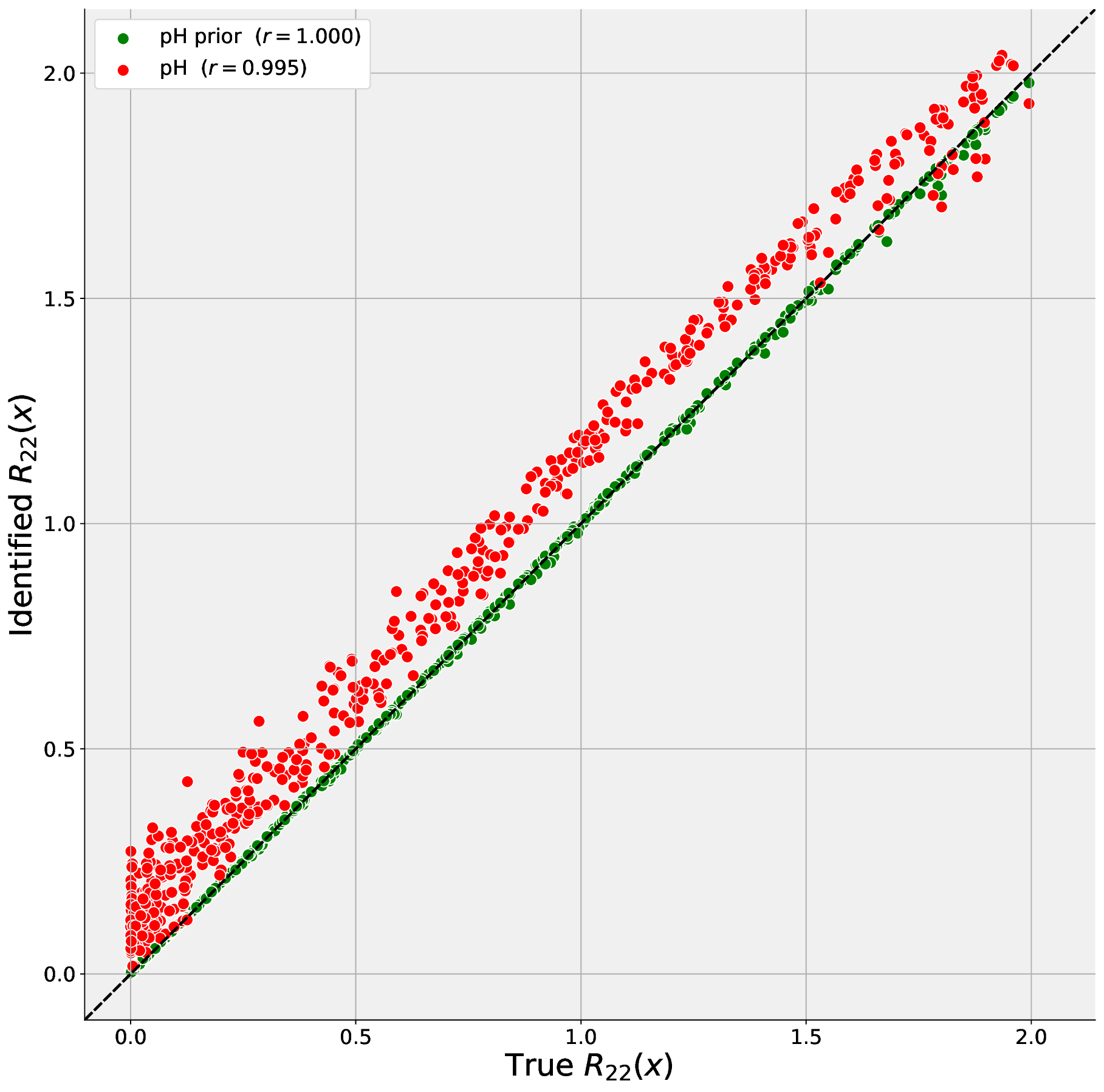}
        \caption{Diagonal entry $R_{22}$ of $R$}
        \label{fig:Compare_R}
    \end{subfigure}
    
    \caption{Scatter plot to visualize the correlation between identified and ground-truth quantities. The corresponding $x$-values are taken from the first $20$ trajectories of the validation dataset.}
    \label{fig:Scatter}
\end{figure}

An interesting question is how well our method performs in identifying the true system parameters. One cannot expect to fully recover the system's coefficients since one could, for example, add a constant to the Hamiltonian or move a scaling factor from the $J(x) - R(x)$ term to the Hamiltonian. However, in practice, we observed strong correlations between the identified parameters and the true parameters, typically even more if one uses prior knowledge. 
To demonstrate this we consider scatter plots. Mathematically these represent the relation $\{(h(z), g(z))\, |\, z \in S\}$ for some functions $h,g : S \rightarrow \mathbb{R}$ and  some set $S$. One can represent these graphically by evaluating the two functions on a point cloud and plotting the values $(h(z), g(z))$ in the plane. We use scatter plots as shown in Figure~\ref{fig:Scatter} to visualize the correlation between identified and true system components.
We consider non-constant entries of the system matrices or the gradient of the Hamiltonian. A~cumulation of the resulting points on a straight line would mean that the two observables are correlated. In addition, if the line represents the identity function with a slope of 1, every identified value is equal to the true value. Thus, slope values closer to 1 result in better identification of the true system parameters.
For example in Figure~\ref{fig:Compare_H}, we can see the correlation between the identified Hamiltonian component (red dots) and the true one (diagonal line). We can see that the red dots tend to accumulate around a straight line which indicates a very high correlation with the true system component. This is also the case for the function $R_{22}$ as we can see in Figure~\ref{fig:Compare_R}. We can moreover see in Figure~\ref{fig:Scatter} that including a prior leads to a significantly greater alignment between the identified entries of $R$ (green dots) and the true values of $R$. In this case, we can even see that the green dots do not merely accumulate around a straight line, but even around the diagonal line which indicates that the true system component has been accurately identified. This shows that adding a prior can help to tackle the problem that a port-Hamiltonian system can have many different port-Hamiltonian reparametrizations.
Note that we do not use a prior on $H$ for this experiment as highlighted in Table~\ref{tab:params_scatter}, so the Hamiltonian needs to be identified using a neural network in Figure~\ref{fig:Compare_H} for the prior variant.

We conclude that our framework can produce identified components of the system with high correlation to the true system components. Moreover, the correlation can be even higher when the right prior is incorporated to other components known in advance.

\begin{table}[h!]
\centering
\resizebox{\textwidth}{!}{
\begin{tabular}{|c |c |c |c |c |c |c |c |c |c| c| c|}
%\toprule
\hline
arch. & $N_{\mathrm{tra}}$& ep. & $\gamma$ &depth &dim & $J$ & $R$ & $H$ & $B$ & SNR &NMAE\\
\hline\hline
pH & $32$& $4000$ & $0.25$ & $3$ &$16$ & MLP & MLP & MLP & MLP & na &$0.0089$\\
\hline
pH prior & $32$& $4000$ & $0.25$ & $3$ &$16$ & const. & MLP & MLP & const.& na&$0.023$\\
\hline
%\bottomrule
\end{tabular}
}
\caption{Training parameters used in Figure~\ref{fig:Scatter}.}
\label{tab:params_scatter}
\end{table}

\subsection{Magnetically levitated ball}
In this example from \cite{Beckers22}, we consider the model of an iron ball of mass $m$ in the magnetic field of a controlled inductor, where the latter incorporates active control terms to stabilize the system's inherent open-loop instability. A more detailed explanation can be found in \cite{Beckers22}. The states of this system are the vertical position of the ball $x_1$, its momentum $x_2$ and the magnetic flux $x_3$ and it can be written in the form 
\[
\dot x=\bigg(\underbrace{\begin{bmatrix}
0 & 1 & 0 \\
-1 & 0 & 0 \\
0 & 0 & 0
\end{bmatrix}}_{=J}-\underbrace{\begin{bmatrix}
0 & 0 & 0 \\
0 & c \cdot |x_2| & 0 \\
0 & 0 & \frac{1}{R_{\mathrm{el}}}
\end{bmatrix}}_{=R(x)}\bigg)\nabla H(x)+\underbrace{\begin{bmatrix}
    0\\0 \\ 1
\end{bmatrix}}_{=:B}u(t)
\]
with the non-quadratic Hamiltonian that is given by 
\[
H(x) =\frac{1}{2m}x_2^2+\frac12\frac{x_3^2}{L(x_1)}.
\] 
For simulations, we set the model parameters as listed in Table~\ref{tab:ironball}.
%and set $m = 0.1$ for the mass of the ball. 

\begin{figure}[H]
    \centering
    \includegraphics[width=1\textwidth]{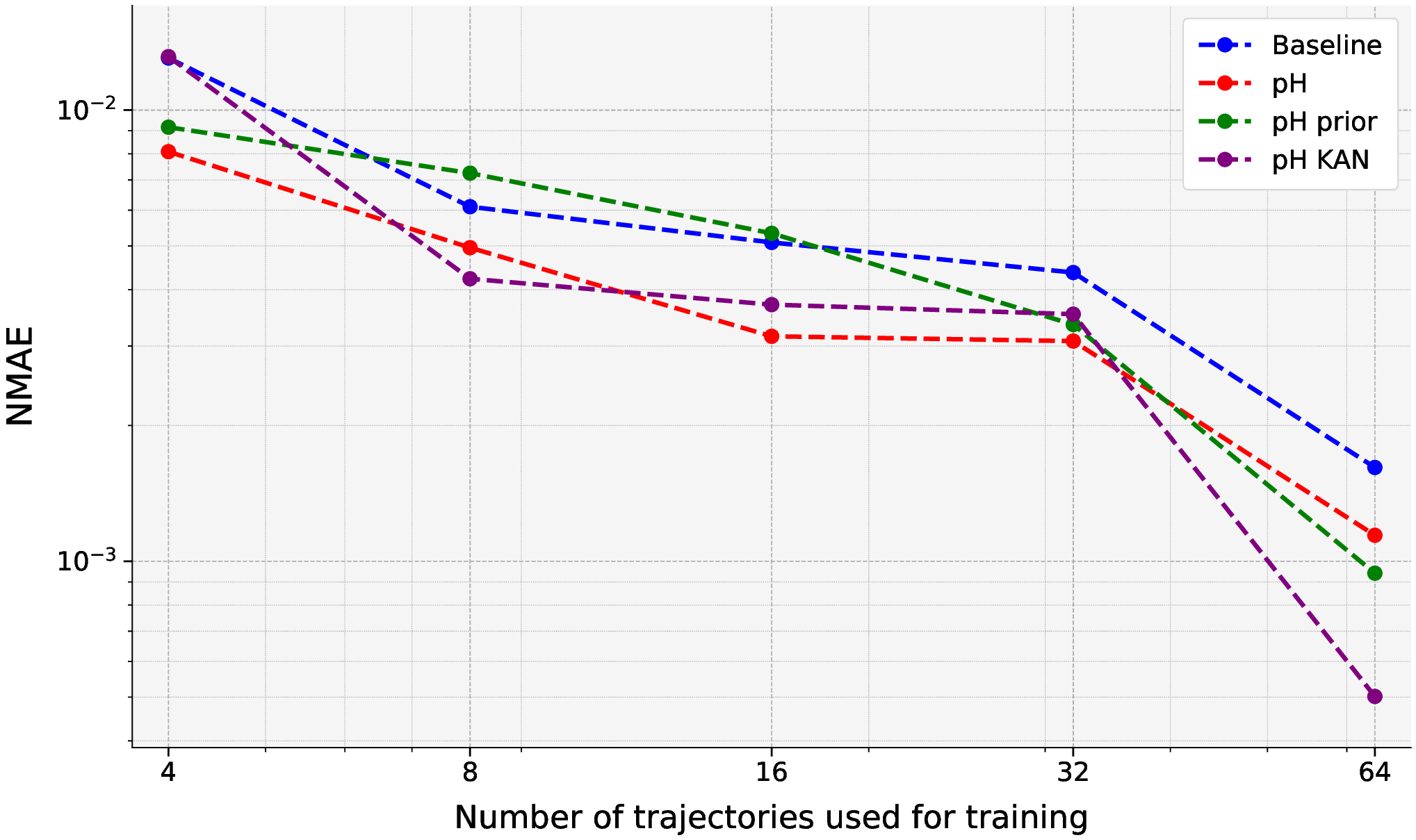}
    \caption{Average validation error for the ball example as a function of varying quantities of trajectories employed in training a baseline model, a default pH setup, the KAN implementation and a pH setup incorporating prior knowledge.}
    \label{fig:Scaling Ball}
\end{figure}

\begin{table}[H]
\centering
\begin{tabular}{|c |c |c |c |}
%\toprule
\hline
symbol& meaning & value & unit \\
\hline\hline
$m$ & ball mass & 0.1 & $[{\rm kg}]$ \\  
$L(x_1)$ & inductance at height $x_1$ & $(0.1+x_1^2)^{-1}$ & $[{\rm H}]$ 
\\
$R_{\mathrm{el}}$ & electrical resistance  & 0.1 & $[\Omega]$ \\
$c$ & drag coefficient  & 1 & -\\
\hline
%\bottomrule
\end{tabular}
\caption{Iron Ball System Parameters}
\label{tab:ironball}
\end{table}
The amplitude of the input is set to $a = 1$ to rescale the signal and we sample the initial states uniformly from $\begin{pmatrix}
x_{1,0}, x_{2,0}, x_{3,0}
\end{pmatrix} \sim [-0.5, 2] \times [-0.2, 0.2] \times [-3, 5]$.
As in the previous example, we consider the comparison between a baseline without any physical knowledge, our default pH model, a model with $J, R, H, B$ parametrized using KANs and a pH model with prior knowledge about which parts of the model are nonlinear. The results for different numbers of trajectories in the training set are shown in Figure~\ref{fig:Scaling Ball}. We notice that the validation error decreases as the number of trajectories increases.

The plot in Figure~\ref{fig:Scaling Ball} indicates that our default pH implementation slightly outperforms the baseline. This time, the incorporation of prior knowledge does not yield a significant performance improvement. Furthermore, the KAN-based approach achieves about the same level of accuracy compared to the MLP-based approach. 

\begin{table}[h!]
\centering
\resizebox{\textwidth}{!}{%
\begin{tabular}{|c |c |c |c |c |c |c |c |c |c| c| }
%\toprule
\hline
arch. & $N_{\mathrm{tra}}$& ep. & $\gamma$ &depth &dim & $J$ & $R$ & $H$ & $B$ & SNR\\
\hline\hline
pH & $4,8,16,32,64$& $500, 1000, 2000, 4000, 8000$ & $0.25$  &$3$&$4, 6, \dots, 26$ & MLP & MLP & MLP & MLP &na\\
\hline
pH KAN & $4,8,16,32, 64$& $500, 1000, 2000, 4000, 8000$ & $0.25$  &$3$&$2, 3, 4, 5, 6, 7$ & KAN & KAN & KAN & KAN& na\\
\hline
pH prior & $4,8,16,32,64$& $500, 1000, 2000, 4000, 8000$ & $0.25$ &$3$ &$4, 6, \dots, 26$ & const. & MLP & MLP & const. & na\\
\hline
baseline & $4,8,16,32,64$& $500, 1000, 2000, 4000, 8000$ & $0.25$ &$3$ &$4, 6, \dots, 26$ & na & na & na & na &na\\
\hline
%\bottomrule
\end{tabular}
}
\caption{Training parameters used in Figure~\ref{fig:Scaling Ball} for the ball example.}
\label{tab:params_scaling_ball}
\end{table}

In order to investigate if the chosen neural network architecture and training configuration is the right one, we inspect the behaviour of a single state trajectory for models trained with $32$ trajectories.

\begin{figure}[H]
\hspace{-1.5cm}
    \includegraphics[scale=0.22]{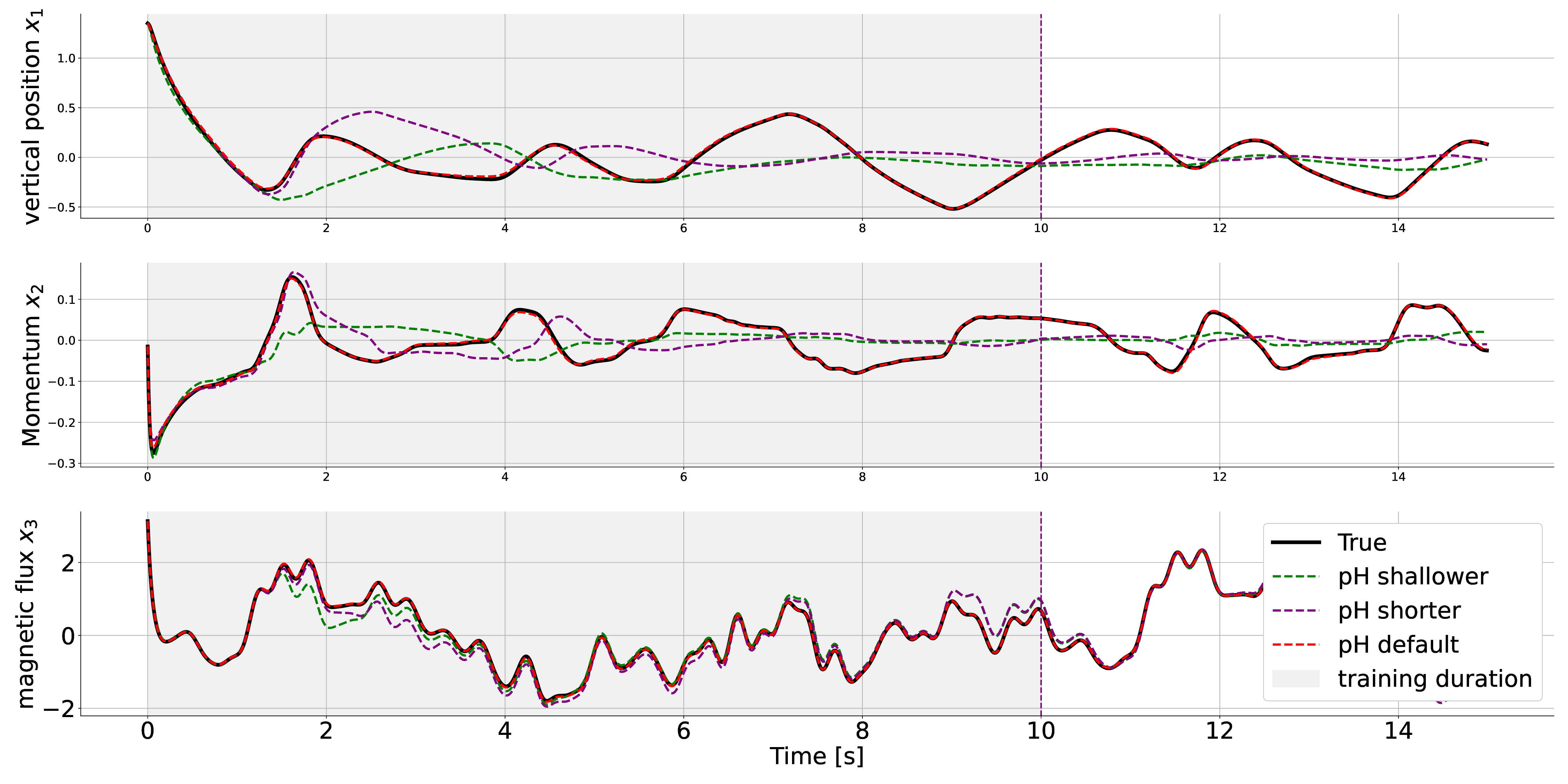}
    \caption{The trajectory of the ball for different pH training setups. The vertical line at $10\,\mathrm{s}$ indicates the final time used in the training.
    }
    \label{fig:Recipe}
\end{figure}

A resulting trajectory of the default pH model is shown in Figure~\ref{fig:Recipe} and compared to the true trajectory. In addition, the same figure shows two other configurations. The first is a shallow network with only one hidden layer. The other is our default pH configuration with a shorter training time of $20$ times fewer epochs. We can see that the latter two configurations deviate more significantly from the ground truth trajectory. This can also be verified by inspecting the NMAE-scores in Table~\ref{tab:params_recipe}. It becomes evident that both the training time and the network depth are crucial hyperparameters for our method.

\begin{table}[h]
\centering
\resizebox{\textwidth}{!}{
\begin{tabular}{|c |c |c |c |c |c |c |c |c |c| c| c|}
%\toprule
\hline
arch. & $N_{\mathrm{tra}}$& ep. & $\gamma$ &depth &dim & $J$ & $R$ & $H$ & $B$ & SNR&NMAE\\
\hline\hline
pH default & $32$& $4000$ & $0.25$  & $3$ &$16$ & MLP & MLP & MLP & MLP &na& $0.0039$\\
\hline
pH shorter & $32$& $200$ & $0.25$  &$3$ & $16$ & const. & MLP & MLP & const.&na&$0.011$\\
\hline
pH shallower & $32$& $4000$ & $0.25$  &$1$ &$16$ & const. & MLP & MLP & const.&na&$0.016$\\
\hline
%\bottomrule
\end{tabular}
}
\caption{Training parameters used in Figure~\ref{fig:Recipe} for the ball example.}
\label{tab:params_recipe}
\end{table}

In Table~\ref{tab:beta}, we ablate on the influence of the hyperparameter $\gamma$ with $32$ trajectories on the ball dataset. We can see that $\gamma \in \{0.25, 0.5\}$ yield approximately the same result while $\gamma\in\{0, 1\}$ yield slightly worse results. Moreover, $\gamma=0$ would mean that the model is not able to accurately reproduce the correct output, because it has no influence on the loss function anymore.
\begin{table}[h!]
\centering
\resizebox{\textwidth}{!}{
\begin{tabular}{|c |c |c |c |c |c |c |c |c |c| c| c|}
%\toprule
\hline
arch. & $N_{\mathrm{tra}}$& ep. & $\gamma$ &depth &dim & $J$ & $R$ & $H$ & $B$ & SNR&NMAE\\
\hline\hline
pH default & $32$& $4000$ & $0$  & $3$ &$16$ & MLP & MLP & MLP & MLP &na& $0.0056$\\
\hline
pH default & $32$& $4000$ & $0.25$  & $3$ &$16$ & MLP & MLP & MLP & MLP &na& $0.0039$\\
\hline
pH default & $32$& $4000$ & $0.5$  & $3$ &$16$ & MLP & MLP & MLP & MLP &na& $0.0038$\\
\hline
pH default & $32$& $4000$ & $1$  & $3$ &$16$ & MLP & MLP & MLP & MLP &na& $0.0046$\\
\hline
%\bottomrule
\end{tabular}
}
\caption{Training parameters for different values of $\gamma$ for the ball example.}
\label{tab:beta}
\end{table}

\subsection{Permanent Magnet Synchronous Motor (PMSM)}
Consider the nonlinear dynamics of a 3-phase Permanent Magnet Synchronous Motor (PMSM) in the $dq$-coordinates which is described in \cite{Spirito2023,Vu23} using the pH formulation as follows: 
\[
\begin{bmatrix}
\dot{\varphi}_d \\
\dot{\varphi}_q \\
\dot{p}
\end{bmatrix} \hspace{-3pt}
= \hspace{-3pt}
\bigg(
\underbrace{\begin{bmatrix}
0 & 0 & \hspace{-12pt}\varphi_q \\
0 & 0 & \hspace{-12pt}-\varphi_d - \Phi_{pm} \\
-\varphi_q & \varphi_d + \Phi_{pm} & \hspace{-12pt}0
\end{bmatrix}}_{=J(x)}
-
\underbrace{\begin{bmatrix}
r & 0 & 0 \\
0 & r & 0 \\
0 & 0 & \beta
\end{bmatrix}}_{=R}
\bigg)
\underbrace{\begin{bmatrix}
\frac{\varphi_d}{L} \\
\frac{\varphi_q}{L} \\
\frac{p}{J_m}
\end{bmatrix}}_{=\nabla H(x)}
+
\underbrace{\begin{bmatrix}
1 & 0 \\
0 & 1 \\
0 & 0
\end{bmatrix}}_{=B}
\begin{bmatrix}
v_d \\
v_q
\end{bmatrix},
\]
where the state vector $x$ is composed of the stator magnetic fluxes $\varphi_d$, $\varphi_q$, and the rotor momentum $p$. Moreover, $\Phi_{pm}$ is the permanent magnet flux of the rotor, $L$ is the phase inductance, $r$ is the stator resistance, $J_m$ is the rotor moment of inertia, and $\beta\geq 0$ is the viscous friction coefficient. Furthermore, the control inputs $v_d$ and $v_q$ are the stator voltages and the  Hamiltonian is a quadratic function that is given by 
\[
H(x) = \frac{\varphi_d^2}{2L} + \frac{\varphi_q^2}{2L} + \frac{p^2}{2J_m}.
\]

To obtain the simulation data for the learning-based system identification task, we choose the model parameters that are listed in Table~\ref{tab:PMSM}. 
\begin{table}[h!]
\centering
\begin{tabular}{|c |c |c |c |}
%\toprule
\hline
symbol& meaning & value & unit \\
\hline\hline
$J_m$ & inertia & 0.012 & $[{\rm kg \cdot m^2}]$ \\   $L$ & phase inductance & $3.8 \cdot 10^{-3}$ & $[{\rm H}]$ \\ 
$\beta$ & viscous friction coefficient & 0.0026 & $[{\rm Nms/rad}]$ \\ $r$ & phase resistance & 0.225 & $[\Omega]$ \\
$\Phi_{pm}$ & constant rotor magnetic flux & 0.17 & $[{\rm Wb}]$ \\
\hline
%\bottomrule
\end{tabular}
\caption{PMSM System Parameters}
\label{tab:PMSM}
\end{table}\\
For the amplitude of the input, we use $a = 5$ to scale the signal and sample the initial conditions uniformly from $\begin{pmatrix}
    \varphi_{d,0}, \varphi_{q,0}, p_0
\end{pmatrix} \sim [-0.5, 0.5]^2 \times [-1, 1]$.

\begin{figure}[H]
    \centering
    \includegraphics[width=1\textwidth]{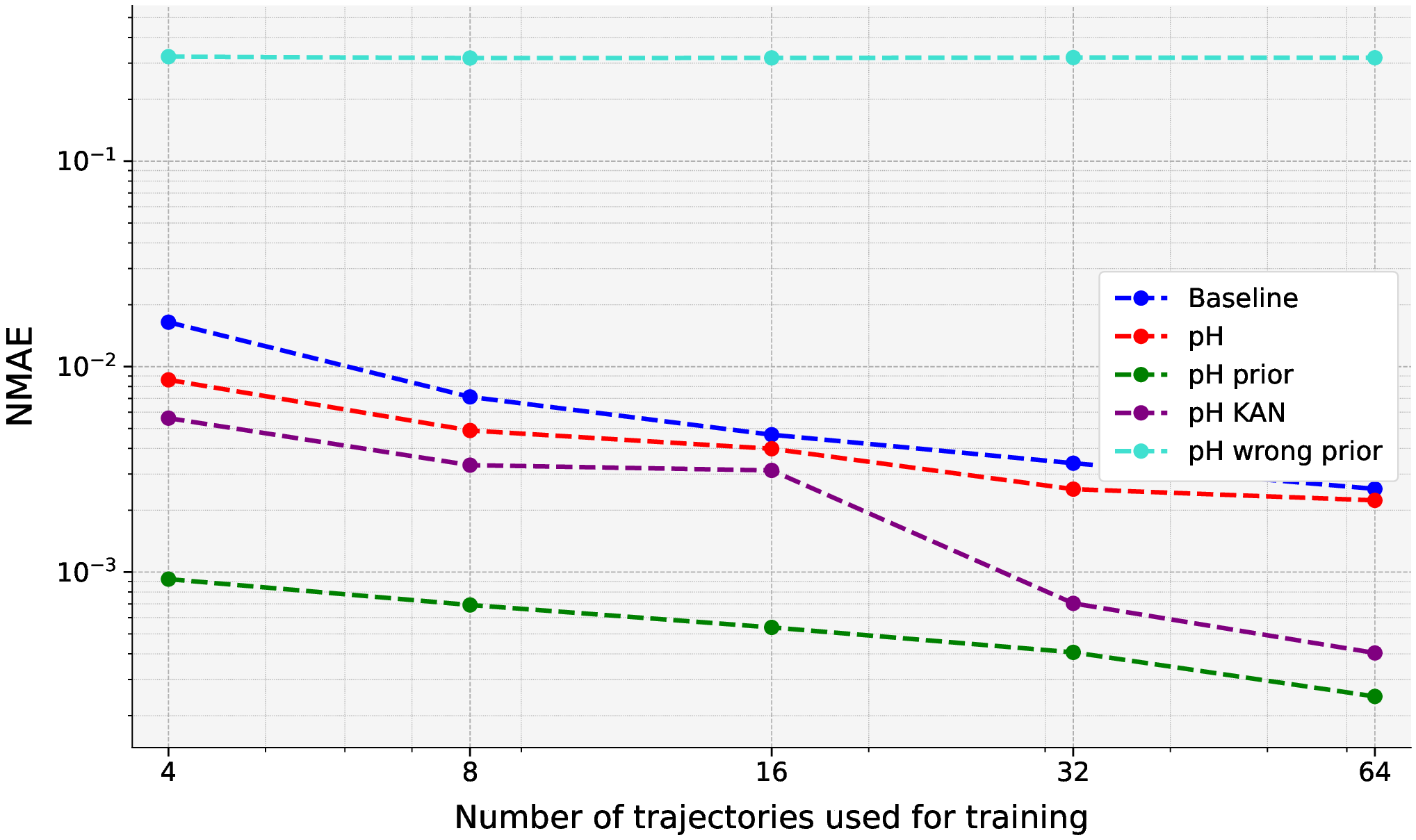}
    \caption{Average validation error for the motor example as a function of varying quantities of trajectories employed in training a baseline model, a default pH setup, a KAN implementation and a pH setup incorporating prior knowledge as well as a setup incorporating a wrong prior.}
    \label{fig:Scaling_motor}
\end{figure}

In this example, $J$ is the state-dependent part of the pH model. We can see in Figure~\ref{fig:Scaling_motor} that the pH model outperforms the baseline. In addition, if we include the prior information that $R$ and $B$ are constant and the Hamiltonian is quadratic, we obtain significantly better results.
Contrary to our findings in the previous examples, this time the KAN-based approach significantly outperforms the MLP-based approach. As discussed in the mass-spring-damper example, this seems to be frequency-related. This indicates that there seems to be no universally best architecture. Instead, it is necessary to try out the different options (MLP, KAN and prior if available). To show a potential downside of using a prior, we added a model with a wrong prior knowledge where the nonlinear component $J$ is also assumed to be linear. We can see that this model with a wrong prior underperforms significantly compared to other variants and does not even improve with more data.

\begin{table}[h!]
\centering
\resizebox{\textwidth}{!}{%
\begin{tabular}{|c |c |c |c |c |c |c |c |c |c| c| }
%\toprule
\hline
arch. & $N_{\mathrm{tra}}$& ep. & $\gamma$ &depth &dim & $J$ & $R$ & $H$ & $B$ & SNR\\
\hline\hline
pH & $4,8,16,32,64$& $500, 1000, 2000, 4000, 8000$ & $0.25$  &$3$&$4, 6, \dots, 26$ & MLP & MLP & MLP & MLP &na\\
\hline
pH KAN & $4,8,16,32, 64$& $500, 1000, 2000, 4000, 8000$ & $0.25$  &$3$&$2, 3, 4, 5, 6, 7$ & KAN & KAN & KAN & KAN& na\\
\hline
pH prior & $4,8,16,32,64$& $500, 1000, 2000, 4000, 8000$ & $0.25$ &$3$ &$4, 6, \dots, 26$ & MLP & const. & quadr. & const. & na\\
\hline
pH wrong prior & $4,8,16,32,64$& $500, 1000, 2000, 4000, 8000$ & $0.25$ &$3$ &na & const. & const. & quadr. & const. & na\\
\hline
baseline & $4,8,16,32,64$& $500, 1000, 2000, 4000, 8000$ & $0.25$ &$3$ &$4, 6, \dots, 26$ & na & na & na & na &na\\
\hline
%\bottomrule
\end{tabular}
}
\caption{Training parameters used in Figure~\ref{fig:Scaling_motor}.}
\label{tab:params_scaling_motor}
\end{table}

\begin{figure} [H]
\centering
\includegraphics[width=1\linewidth]{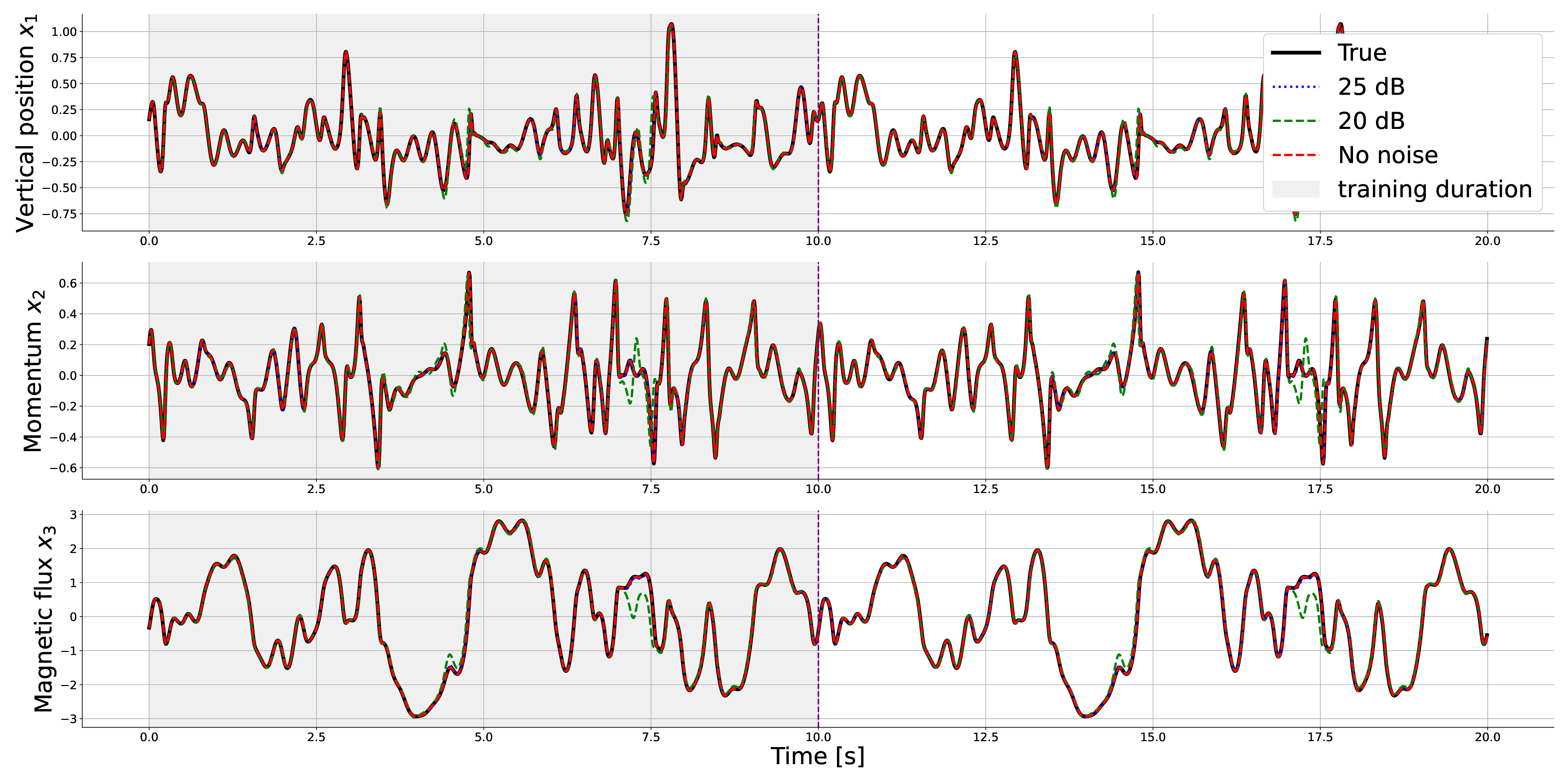}
\caption{State trajectories of the  motor example based on the pH model with different SNR levels in the training data.}
    \label{fig:noise}
\end{figure}

Furthermore, for this example, we consider a preliminary study on noisy data. Uniform white noise was added to the inputs and outputs during training, while the states and derivatives of these were kept noise-free. Different signal-to-noise ratio (SNR) levels are compared in Table~\ref{tab:noise}. We can see that the NMAE increases gradually with more noise. Moreover for $\rm SNR = 35\, \mathrm{dB}$ the NMAE is similar to the one of the model trained without noise. In Figure~\ref{fig:noise} we can see the predictions of the model without noise and the ones with $\rm SNR = 20, 25 \, \mathrm{dB}$ for one trajectory. We can see that the model without noise and the model with $\rm SNR = 25\, \mathrm{dB}$ are very close to the ground truth trajectory while the model with $\rm SNR = 20\, \mathrm{dB}$ can deviate significantly from it. We do not show  $\rm SNR = 30, 35 \, \mathrm{dB}$ levels as they are indistinguishable from the noiseless case. In this experiment, we showed that our method remains quite robust to noise, although this problem is not tackled specifically.

\begin{table}[h]
\centering
\resizebox{\textwidth}{!}{%
\begin{tabular}{|c |c |c |c |c |c |c| |c |c |c |c| c| c|}
%\toprule
\hline
arch. & $N_{\mathrm{tra}}$& ep. & $\gamma$ & wd &depth &dim & $J$ & $R$ & $H$ & $B$ & SNR&NMAE\\
\hline\hline
$20 \mathrm{dB}$& $16$& $4000$ & $0.25$ & $0.01$ &$3$&$16$ &  MLP & MLP & MLP & MLP&$20\mathrm{dB}$ &$0.095$\\
\hline
$25 \mathrm{dB}$& $16$& $4000$ & $0.25$ & $0.01$ &$3$&$16$ &  MLP & MLP & MLP & MLP&$25\mathrm{dB}$ &$0.045$\\
\hline
$30\mathrm{dB}$ & $16$& $4000$ & $0.25$ & $0.01$&$3$ &$16$ &  MLP & MLP & MLP & MLP &$30\mathrm{dB}$&$0.019$\\
\hline
$35\mathrm{dB}$ & $16$& $4000$ & $0.25$ & $0.01$ &$3$&$16$ &  MLP & MLP & MLP & MLP&$35\mathrm{dB}$&$0.010$\\
\hline
no Noise & $16$& $4000$ & $0.25$ & $0.01$ &$3$&$16$ & MLP & MLP & MLP & MLP &na&$0.0065$\\
\hline
%\bottomrule
\end{tabular}
}
\caption{Training parameters used to study the noise tolerance of our method.}
\label{tab:noise}
\end{table}

As a summary, from our experiments, we noticed that enforcing the port-Hamiltonian framework, generally, does not deteriorate the accuracy and sometimes even improves the results for the system identification of nonlinear systems. If additional information about the model is available, such as the linearity of certain components, it can be integrated into the framework, leading to a significant reduction in validation error. We have also seen that for some systems the KAN-based approach can yield even better results than the MLP-based identification.

\section*{Code and Data Availability}

The code used to replicate all numerical experiments presented in the paper is available at \url{https://github.com/trawler0/ISO-pHNN}.

\section{Conclusion}

We proposed a framework called ISO-pHNN for the identification of system matrices for nonlinear port-Hamiltonian systems utilizing input-state-output data. The data is used to approximate each system matrix $J$, $R$ and $B$ as well as the Hamiltonian $H$ with a multilayer perceptron (MLP), a Kolmogorov-Arnold network (KAN), or, whenever prior information is available, a weighted combination of ansatz functions. The weights are subsequently optimized using the provided data. The effectiveness of our approach is demonstrated through several examples that exhibit nonlinearities in skew-symmetric components, dissipative terms, or the Hamiltonian. 
Our method can recover the true system components, particularly well when prior information is used.
Future research will explore applications to more extensive examples, potentially derived from discretized partial differential equations, and will focus on identification from solely input-output data that may be incomplete. In addition, future work will focus on further enhancing the practical applicability of ISO-pHNN by removing the reliance on state derivatives during training and by adopting uncertainty quantification (UQ) techniques that evaluate results across multiple independent runs and report confidence intervals.

  \bibliographystyle{abbrv} 
  \bibliography{sample}

\begin{thebibliography}{10}

\bibitem{EffKAN24}
\url{https://github.com/Blealtan/efficient-kan}, 2024.

\bibitem{Beckers22}
T.~Beckers, J.~Seidman, P.~Perdikaris, and G.~J. Pappas.
\newblock Gaussian process port-{H}amiltonian systems: Bayesian learning with
  physics prior.
\newblock In {\em IEEE 61st CDC}, pages 1447--1453, 2022.

\bibitem{Benner2021}
P.~Benner, P.~Goyal, and P.~{Van Dooren}.
\newblock Identification of port-{H}amiltonian systems from frequency response
  data.
\newblock {\em Systems Control Lett.}, 143:104741, 2020.

\bibitem{Brunton_2016}
S.~L. Brunton, J.~L. Proctor, and J.~N. Kutz.
\newblock Discovering governing equations from data by sparse identification of
  nonlinear dynamical systems.
\newblock {\em Proc. Natl. Acad. Sci. U.S.A.}, 113(15):3932–3937, 2016.

\bibitem{Car24}
F.~L. Cardoso-Ribeiro, G.~Haine, Y.~{Le Gorrec}, D.~Matignon, and H.~Ramirez.
\newblock Port-{H}amiltonian formulations for the modeling, simulation and
  control of fluids.
\newblock {\em Comput. \& Fluids}, 283:106407, 2024.

\bibitem{Chen2018}
R.~T.~Q. Chen, Y.~Rubanova, J.~Bettencourt, and D.~K. Duvenaud.
\newblock Neural ordinary differential equations.
\newblock In S.~Bengio, H.~Wallach, H.~Larochelle, K.~Grauman, N.~Cesa-Bianchi,
  and R.~Garnett, editors, {\em NeurIPS}, volume~31. Curran Associates, Inc.,
  2018.

\bibitem{Cherifi2020}
K.~Cherifi.
\newblock An overview on recent machine learning techniques for port
  {H}amiltonian systems.
\newblock {\em Phys. D}, 411:132620, 2020.

\bibitem{CheB21}
K.~Cherifi and A.~Brugnoli.
\newblock Application of data-driven realizations to port-{H}amiltonian
  flexible structures.
\newblock {\em IFAC-PapersOnLine}, 54(19):180--185, 2021.

\bibitem{CheGH23}
K.~Cherifi, H.~Gernandt, and D.~Hinsen.
\newblock The difference between port-{Hamiltonian}, passive and positive real
  descriptor systems.
\newblock {\em Math. Control Signals Systems}, 36:451--482, 2023.

\bibitem{Che22}
K.~Cherifi, P.~Goyal, and P.~Benner.
\newblock A non-intrusive method to inferring linear port-{H}amiltonian
  realizations using time-domain data.
\newblock {\em ETNA : Special Issue SciML}, 56:102 -- 116, 2022.

\bibitem{Cherifi2019}
K.~Cherifi, V.~Mehrmann, and K.~Hariche.
\newblock Numerical methods to compute a minimal realization of a
  port-{H}amiltonian system.
\newblock {\em Preprint arXiv:1903.07042}, 2019.

\bibitem{Cybenko1989}
G.~Cybenko.
\newblock Approximation by superpositions of a sigmoidal function.
\newblock {\em Math. Control Signals Systems}, 2:303--314, 1989.

\bibitem{Desai2021}
S.~A. Desai, M.~Mattheakis, D.~Sondak, P.~Protopapas, and S.~J. Roberts.
\newblock Port-{H}amiltonian neural networks for learning explicit
  time-dependent dynamical systems.
\newblock {\em Phys. Rev. E}, 104:034312, Sep 2021.

\bibitem{dormand1980family}
J.~R. Dormand and P.~J. Prince.
\newblock A family of embedded {R}unge-{K}utta formulae.
\newblock {\em J. Comput. Appl. Math.}, 6(1):19--26, 1980.

\bibitem{Dan24}
H.~Dänschel, L.~Lentz, and U.~von Wagner.
\newblock Error measures and solution artifacts of the harmonic balance method
  on the example of the softening {D}uffing oscillator.
\newblock {\em J. Theor. Appl. Mech.}, 62(2):435--455, 2024.

\bibitem{Eidnes23}
S.~Eidnes, A.~Stasik, C.~Sterud, E.~Bøhn, and S.~Riemer-Sørensen.
\newblock Pseudo-{H}amiltonian neural networks with state-dependent external
  forces.
\newblock {\em Phys. D}, 446:133673, 2023.

\bibitem{elfwing2018sigmoid}
S.~Elfwing, E.~Uchibe, and K.~Doya.
\newblock Sigmoid-weighted linear units for neural network function
  approximation in reinforcement learning.
\newblock {\em Neural Networks}, 107:3--11, 2018.

\bibitem{Finzi2020}
M.~Finzi, K.~A. Wang, and A.~G. Wilson.
\newblock Simplifying {H}amiltonian and {L}agrangian neural networks via
  explicit constraints.
\newblock In H.~Larochelle, M.~Ranzato, R.~Hadsell, M.~Balcan, and H.~Lin,
  editors, {\em NeurIPS}, volume~33, pages 13880--13889. Curran Associates,
  Inc., 2020.

\bibitem{GerHRS21}
H.~Gernandt, F.~Haller, T.~Reis, and A.~van~der Schaft.
\newblock Port-{H}amiltonian formulation of nonlinear electrical circuits.
\newblock {\em J.\ Geom.\ Phys.}, 159:103959, 2021.

\bibitem{GerSZMS24}
H.~Gernandt, B.~Severino, X.~Zhang, V.~Mehrmann, and K.~Strunz.
\newblock Port-{H}amiltonian modeling and control of electric vehicle charging
  stations.
\newblock {\em IEEE Trans. Transp. Electrific.}, 2024.

\bibitem{GilS18}
N.~Gillis and P.~Sharma.
\newblock Finding the nearest positive-real system.
\newblock {\em SIAM J. Numer. Anal.}, 56(2):1022--1047, 2018.

\bibitem{GorH24}
N.~A. G\"{o}ring, F.~Hess, M.~Brenner, Z.~Monfared, and D.~Durstewitz.
\newblock Out-of-domain generalization in dynamical systems reconstruction.
\newblock In {\em ICML}, volume 235, pages 16071--16114. PMLR, 2024.

\bibitem{goyal2018accuratelargeminibatchsgd}
P.~Goyal, P.~Dollár, R.~Girshick, P.~Noordhuis, L.~Wesolowski, A.~Kyrola,
  A.~Tulloch, Y.~Jia, and K.~He.
\newblock Accurate, large minibatch {SGD}: Training {I}mage{N}et in 1 hour.
\newblock {\em Preprint arXiv:1706.02677}, 2018.

\bibitem{Greydanus2019}
S.~Greydanus, M.~Dzamba, and J.~Yosinski.
\newblock Hamiltonian neural networks.
\newblock In H.~Wallach, H.~Larochelle, A.~Beygelzimer, F.~d\textquotesingle
  Alch\'{e}-Buc, E.~Fox, and R.~Garnett, editors, {\em NeurIPS}, volume~32.
  Curran Associates, Inc., 2019.

\bibitem{Guen2020}
V.~L. Guen and N.~Thome.
\newblock Disentangling physical dynamics from unknown factors for unsupervised
  video prediction.
\newblock In {\em 2020 IEEE/CVF CVPR}, pages 11471--11481, 2020.

\bibitem{Gunther2023}
M.~Günther, B.~Jacob, and C.~Totzeck.
\newblock Data-driven adjoint-based calibration of port-{H}amiltonian systems
  in time domain.
\newblock {\em Math. Control Signals Systems}, 36:957–977, 2024.

\bibitem{Gunther2024}
M.~Günther, B.~Jacob, and C.~Totzeck.
\newblock Structure-preserving identification of port-{H}amiltonian systems—a
  sensitivity-based approach.
\newblock In {\em Scientific Computing in Electrical Engineering, M. van
  Beurden et al., ed.}, pages 167--174, Cham, 2024. Springer Nature
  Switzerland.

\bibitem{Hastie2009}
T.~Hastie, R.~Tibshirani, and J.~Friedman.
\newblock {\em Neural Networks}, pages 389--416.
\newblock Springer NY, 2009.

\bibitem{Hornik1989}
K.~Hornik, M.~Stinchcombe, and H.~White.
\newblock Multilayer feedforward networks are universal approximators.
\newblock {\em Neural Networks}, 2(5):359--366, 1989.

\bibitem{JacT24}
B.~Jacob and C.~Totzeck.
\newblock Port-{H}amiltonian structure of interacting particle systems and its
  mean-field limit.
\newblock {\em Multiscale Model. Simul.}, 22(4):1247--1266, 2024.

\bibitem{kingma2014adam}
D.~P. Kingma and J.~Ba.
\newblock Adam: A method for stochastic optimization.
\newblock {\em Preprint arXiv:1412.6980}, 2014.

\bibitem{Koenig2024KANODE}
B.~C. Koenig, S.~Kim, and S.~Deng.
\newblock {KAN}-{ODE}s: Kolmogorov--{A}rnold network ordinary differential
  equations for learning dynamical systems and hidden physics.
\newblock {\em Comput. Methods Appl. Mech. Engrg.}, 432:117397, 2024.

\bibitem{lei2016layer}
J.~Lei~Ba, J.~R. Kiros, and G.~E. Hinton.
\newblock Layer normalization.
\newblock {\em Preprint arXiv:1607.06450}, 2016.

\bibitem{linares2026controlled}
M.~Linares, G.~Doras, and T.~H{\'e}lie.
\newblock Controlled oscillation modeling using port-{H}amiltonian neural
  networks.
\newblock {\em arXiv preprint arXiv:2602.15704}, 2026.

\bibitem{Liu2024}
Z.~Liu, Y.~Wang, S.~Vaidya, F.~Ruehle, J.~Halverson, M.~Soljacic, T.~Y. Hou,
  and M.~Tegmark.
\newblock {KAN}: Kolmogorov{\textendash}{A}rnold networks.
\newblock In {\em The Thirteenth International Conference on Learning
  Representations}, 2025.

\bibitem{Lohmayer22}
M.~Lohmayer and S.~Leyendecker.
\newblock {EPHS}: A port-{H}amiltonian modelling language.
\newblock {\em IFAC-PapersOnLine}, 55(30):347--352, 2022.

\bibitem{Lopes2015}
N.~Lopes, T.~Hélie, and A.~Falaize.
\newblock Explicit second-order accurate method for the passive guaranteed
  simulation of port-{H}amiltonian systems.
\newblock {\em IFAC-PapersOnLine}, 48(13):223--228, 2015.

\bibitem{loshchilov2017sgdrstochasticgradientdescent}
I.~Loshchilov and F.~Hutter.
\newblock {SGDR}: Stochastic gradient descent with warm restarts.
\newblock In {\em ICLR}, 2017.

\bibitem{Loshchilov2017}
I.~Loshchilov and F.~Hutter.
\newblock Decoupled weight decay regularization.
\newblock In {\em ICLR}, 2019.

\bibitem{Medianu17}
S.~O. Medianu.
\newblock {\em {Identification of Port {H}amiltonian systems}}.
\newblock Theses, {Universit{\'e} Grenoble - Alpes}, 2017.

\bibitem{MehU23}
V.~Mehrmann and B.~Unger.
\newblock Control of port-{H}amiltonian differential-algebraic systems and
  applications.
\newblock {\em Acta Numerica}, pages 395--515, 2023.

\bibitem{OE-PHNN26}
S.~Moradi, G.~I. Beintema, N.~O. Jaensson, R.~Tóth, and M.~Schoukens.
\newblock Port-{H}amiltonian neural networks with output error noise models.
\newblock {\em Automatica}, 187:112892, 2026.

\bibitem{Morandin2023}
R.~Morandin, J.~Nicodemus, and B.~Unger.
\newblock Port-{H}amiltonian dynamic mode decomposition.
\newblock {\em SIAM J. Sci. Comput.}, 45(4):A1690--A1710, 2023.

\bibitem{Najnudel21}
J.~Najnudel, R.~Müller, T.~Hélie, and D.~Roze.
\newblock Identification of nonlinear circuits as port-{H}amiltonian systems.
\newblock In {\em 24th Int. Conf. on Digital Audio Effects (DAFx)}, pages 1--8,
  2021.

\bibitem{Neary2023}
C.~Neary and U.~Topcu.
\newblock Compositional learning of dynamical system models using
  port-{H}amiltonian neural networks.
\newblock In {\em 5th L4DC}, volume 211, pages 1--17, 2023.

\bibitem{paszke2019pytorchimperativestylehighperformance}
A.~Paszke, S.~Gross, F.~Massa, A.~Lerer, J.~Bradbury, G.~Chanan, T.~Killeen,
  Z.~Lin, N.~Gimelshein, L.~Antiga, A.~Desmaison, A.~Kopf, E.~Yang, Z.~DeVito,
  M.~Raison, A.~Tejani, S.~Chilamkurthy, B.~Steiner, L.~Fang, J.~Bai, and
  S.~Chintala.
\newblock Py{T}orch: An imperative style, high-performance deep learning
  library.
\newblock In H.~Wallach, H.~Larochelle, A.~Beygelzimer, F.~d\textquotesingle
  Alch\'{e}-Buc, E.~Fox, and R.~Garnett, editors, {\em NeurIPS}, volume~32.
  Curran Associates, Inc., 2019.

\bibitem{Pfaff2020}
T.~Pfaff, M.~Fortunato, A.~Sanchez-Gonzalez, and P.~Battaglia.
\newblock Learning mesh-based simulation with graph networks.
\newblock In {\em ICLR}, 2021.

\bibitem{Pon24}
C.~Ponce, Y.~Wu, Y.~{Le Gorrec}, and H.~Ramirez.
\newblock A systematic methodology for port-{H}amiltonian modeling of
  multidimensional flexible linear mechanical systems.
\newblock {\em Appl. Math. Model.}, 134:434--451, 2024.

\bibitem{Raissi2019}
M.~Raissi, P.~Perdikaris, and G.~E. Karniadakis.
\newblock Physics-informed neural networks: A deep learning framework for
  solving forward and inverse problems involving nonlinear partial differential
  equations.
\newblock {\em J. Comput. Phys.}, 378:686--707, 2019.

\bibitem{Rettberg2024}
J.~Rettberg, J.~Kneifl, J.~Herb, P.~Buchfink, J.~Fehr, and B.~Haasdonk.
\newblock Data-driven identification of latent port-hamiltonian systems.
\newblock {\em Computational Science and Engineering}, 2(4), 2025.

\bibitem{roschkowski2026neuralscalinglawslearningbased}
M.~Roschkowski, K.~Cherifi, and H.~Gernandt.
\newblock Neural scaling laws for learning-based identification of nonlinear
  systems.
\newblock {\em Preprint arXiv:2512.20447}, 2026.

\bibitem{Schwerdtner2021}
P.~Schwerdtner.
\newblock Port-{H}amiltonian system identification from noisy frequency
  response data.
\newblock {\em Preprint arXiv:2106.11355}, 2021.

\bibitem{Schwerdtner2022}
P.~Schwerdtner, T.~Moser, V.~Mehrmann, and M.~Voigt.
\newblock Optimization-based model order reduction of port-{H}amiltonian
  descriptor systems.
\newblock {\em Systems Control Lett.}, 182:105655, 2023.

\bibitem{SchSBP24}
P.~Schwerdtner, P.~Schulze, J.~Berman, and B.~Peherstorfer.
\newblock Nonlinear embeddings for conserving {H}amiltonians and other
  quantities with {N}eural {G}alerkin schemes.
\newblock {\em SIAM J. Sci. Comput.}, 46(5):C583--C607, 2024.

\bibitem{Sch23}
P.~Schwerdtner and M.~Voigt.
\newblock {SOBMOR}: Structured optimization-based model order reduction.
\newblock {\em SIAM J. Sci. Comput.}, 45(2):A502--A529, 2023.

\bibitem{Spirito2023}
M.~Spirito, Y.~Le~Gorrec, and B.~Maschke.
\newblock Structure-preserving observers for port-{H}amiltonian systems via
  contraction analysis.
\newblock {\em Preprint hal-04344593}, 2023.

\bibitem{Toth2020}
P.~Toth, D.~J. Rezende, A.~Jaegle, S.~Racanière, A.~Botev, and I.~Higgins.
\newblock Hamiltonian generative networks.
\newblock In {\em ICLR}, 2020.

\bibitem{SchJ14}
A.~J. van~der Schaft and D.~Jeltsema.
\newblock Port-{{Hamiltonian Systems Theory}}: {{An Introductory Overview}}.
\newblock {\em Found. Trends Syst. Control}, 1(2-3):173--378, 2014.

\bibitem{vanOMWTJS24}
G.~J.~E. van Otterdijk, S.~Moradi, S.~Weiland, R.~Tóth, N.~O. Jaensson, and
  M.~Schoukens.
\newblock Learning subsystem dynamics in nonlinear systems via
  port-{H}amiltonian neural networks.
\newblock {\em Preprint arXiv:2411.05730}, 2024.

\bibitem{vaswani2023attentionneed}
A.~Vaswani, N.~Shazeer, N.~Parmar, J.~Uszkoreit, L.~Jones, A.~N. Gomez, L.~u.
  Kaiser, and I.~Polosukhin.
\newblock Attention is all you need.
\newblock In I.~Guyon, U.~V. Luxburg, S.~Bengio, H.~Wallach, R.~Fergus,
  S.~Vishwanathan, and R.~Garnett, editors, {\em NeurIPS}, volume~30. Curran
  Associates, Inc., 2017.

\bibitem{virtanen2020scipy}
P.~Virtanen, R.~Gommers, T.~E. Oliphant, M.~Haberland, T.~Reddy, D.~Cournapeau,
  E.~Burovski, P.~Peterson, W.~Weckesser, J.~Bright, et~al.
\newblock Sci{P}y 1.0: fundamental algorithms for scientific computing in
  {P}ython.
\newblock {\em Nat. Methods}, 17(3):261--272, 2020.

\bibitem{Vu23}
N.~M.~T. Vu, T.~H. Pham, I.~Prodan, and L.~Lef{\`{e}}vre.
\newblock Port-{H}amiltonian observer for state-feedback control design.
\newblock In {\em ECC, Bucharest, Romania}, pages 1--6, 2023.

\bibitem{WanSLH25}
Y.~Wang, J.~W. Siegel, Z.~Liu, and T.~Y. Hou.
\newblock On the expressiveness and spectral bias of {KAN}s.
\newblock In {\em ICLR}, 2025.

\bibitem{Yin2021}
Y.~Yin, V.~L. Guen, J.~Dona, E.~de~Bézenac, I.~Ayed, N.~Thome, and
  P.~Gallinari.
\newblock Augmenting physical models with deep networks for complex dynamics
  forecasting.
\newblock {\em J. Stat. Mech. Theory Exp.}, 2021(12):124012, 2021.

\bibitem{Zaspel2024}
P.~Zaspel and M.~Günther.
\newblock Data-driven identification of port-{H}amiltonian {DAE} systems by
  {G}aussian processes.
\newblock {\em Preprint arXiv:2406.18726}, 2024.

\end{thebibliography}

\end{document}